\documentclass[]{aa}

\usepackage[utf8]{inputenc}
\usepackage{graphicx}
\usepackage[varg]{txfonts}

\bibpunct{(}{)}{;}{a}{}{,}
\begin{document}
\title{Searching for HI around MHONGOOSE Galaxies via Spectral Stacking}

\author{S. Veronese\inst{\ref{astron},\ref{kapteyn}} \and W. J. G. de Blok\inst{\ref{astron},\ref{kapteyn},\ref{uct}} \and J. Healy\inst{\ref{astron}} \and D. Kleiner\inst{\ref{astron},\ref{inafca}} \and A. Marasco\inst{\ref{inafpd}} \and F. M. Maccagni\inst{\ref{inafca}} \and P. Kamphuis\inst{\ref{ruhr}} \and E. Brinks\inst{\ref{car}} \and B. W. Holwerda\inst{\ref{ul}} \and N. Zabel\inst{\ref{uct}} \and L. Chemin\inst{\ref{antofagasta}} \and E. A. K. Adams\inst{\ref{astron},\ref{kapteyn}} \and S. Kurapati\inst{\ref{uct},\ref{astron}} \and A. Sorgho\inst{\ref{iaa}} \and K. Spekkens\inst{\ref{kingston}} \and F. Combes\inst{\ref{lerma}} \and D. J. Pisano\inst{\ref{uct}} \and F. Walter\inst{\ref{mpia}} \and P. Amram\inst{\ref{amu}} \and F. Bigiel\inst{\ref{bonn}} \and O. I. Wong\inst{\ref{csiro},\ref{icrar}} \and E. Athanassoula\inst{\ref{amu}}}

\institute{Netherlands Institute for Radio Astronomy (ASTRON), Oude Hoogeveensedijk 4, 7991 PD Dwingeloo, The Netherlands, \email{veronese@astron.nl}\label{astron}
\and Kapteyn Astronomical Institute, University of Groningen, PO Box 800, 9700 AV Groningen, The Netherlands\label{kapteyn}
\and Department of Astronomy, University of Cape Town, Private Bag X3, 7701 Rondebosch, South Africa\label{uct}
\and INAF – Osservatorio Astronomico di Cagliari, Via della Scienza 5, 09047, Selargius, CA, Italy\label{inafca} \and INAF - Padova Astronomical Observatory, Vicolo dell’Osservatorio 5, I-35122 Padova, Italy\label{inafpd} \and Ruhr University Bochum, Faculty of Physics and Astronomy, Astronomical Institute (AIRUB), 44780 Bochum, Germany\label{ruhr} \and Centre for Astrophysics Research, University of Hertfordshire, College Lane, Hatfield AL10 9AB, United Kingdom\label{car} \and Department of Physics and Astronomy, University of Louisville, Natural Science Building 102, 40292 KY Louisville, USA\label{ul} \and Universidad Andr\'es Bello, Facultad de Ciencias Exactas, Departamento de Ciencias Físicas - Instituto de Astrof\'isica, Fern\'andez Concha 700, Las Condes, Santiago, Chile\label{antofagasta} \and Instituto de Astrofısica de Andalucıa (IAA-CSIC), Glorieta de la Astronomıa s/n, E-18008 Granada, Spain\label{iaa} \and Department of Physics, Engineering Physics and Astronomy Queen’s University Kingston, ON K7L 3N6, Canada\label{kingston} \and Observatoire de Paris, LERMA, Collège de France, CNRS, PSL University, Sorbonne University, 75014 Paris, France\label{lerma} \and Dept. of Physics \& Astronomy and Center for Gravitational Waves \& Cosmology, West Virginia University, Morgantown, WV, USA\label{wvu} \and Adjunct Astronomer at Green Bank Observatory, Green Bank, WV, USA\label{gbt} \and Max Planck Institute für Astronomie, Heidelberg, Germany\label{mpia} \and Aix-Marseille Univ., CNRS, CNES, LAM, 38 rue Frédéric Joliot Curie, 13338 Marseille, France\label{amu} \and Argelander-Institut für Astronomie, Auf dem Hügel 71, 53121
Bonn, Germany\label{bonn} \and Australia Telescope National Facility, CSIRO, Space and Astronomy, PO Box 1130, Bentley, WA 6102, Australia\label{csiro} \and International Centre for Radio Astronomy Research, The University of Western Australia, Crawley, WA 6009, Australia\label{icrar}}

\date{Received 2 September 2024 / Accepted 15 November 2024}

\abstract{The observed star formation rates of galaxies in the Local Universe suggests that they are replenishing their gas reservoir across cosmic time. Cosmological simulations predict that this accretion of fresh gas can occur in a hot or a cold mode, yet the existence of low column density ($\sim10^{17}$ cm$^{-2}$) neutral atomic hydrogen (H{\sc i}) tracing the cold mode has not been unambiguously confirmed by observations.\\We present the application of unconstrained spectral stacking to attempt to detect the emission from this H{\sc i} in the Circum-Galactic Medium (CGM) and Inter-Galactic Medium (IGM) of 6 nearby star forming galaxies from the MHONGOOSE sample for which full-depth observations are available.\\Our stacking procedure consists of a standard spectral stacking algorithm coupled with a one-dimensional spectral line finder designed to extract reliable signal close to the noise level.\\In agreement with previous studies, we found that the amount of signal detected outside the H{\sc i} disk is much smaller than implied by simulations. Furthermore, the column density limit that we achieve via stacking ($\sim10^{17}$ cm$^{-2}$) suggests that direct detection of the neutral CGM/IGM component might be challenging in the future, even with the next generation of radio telescopes.}

\keywords{Galaxies: evolution - Galaxies: formation - Radio lines: galaxies - Line: identification}
\maketitle

\section{Introduction}\label{sec:intro}
The process of cold gas accretion, whereby pristine gas from the Inter-Galactic Medium (IGM) \citep{dekel06,sommerville15,danovich15} or processed gas from the Circum-Galactic Medium (CGM) \citep{putman12,afruni21,li21,marasco22} or from galaxy mergers \citep{sancisi08,putman12,blok20,porter22} fuels the growth of galaxies, is a crucial mechanism in galaxy formation and evolution.\\\indent It is believed that neutral atomic hydrogen (H{\sc i}) clouds in the IGM and CGM serve as a tracer for the reservoirs of cool ($T<10^5$ K) gas that can be funnelled onto galaxies, providing the necessary fuel for star formation and driving the growth of stellar mass over cosmic time \citep{sancisi08,madau14,dave17,tumlinson17,chen21,faucher23}.\\\indent The detection and characterization of these H{\sc i} clouds is essential for understanding the efficiency of their accretion onto galaxies \citep{dekel09,putman12,kamphuis22,afruni23,lochhaas23}. By studying their spatial distribution, abundance, and physical properties, we can gain insights into the mechanisms responsible for the transport and deposition of gas from the cosmic web on to galaxies \citep{popping09,sommerville15,eagle,fire2,tng,voort19,ramesh24}. Additionally, the kinematics and gas dynamics of the H{\sc i} clouds can provide valuable information about the accretion processes and the interplay between galaxies and their surrounding environment.\\\indent The presence of $\sim$ kpc-scale, low column density ($N_{HI}\sim10^{17}$ cm$^{-2}$), H{\sc i} clouds in the CGM/IGM has been predicted by simulations \citep{popping09,voort19,ramesh24}, yet not unambiguously confirmed observationally \citep{braun04,wolfe13,wolfe16,tumlinson17,das20,kamphuis22,xu22,liu23,das24}. The detection of CGM/IGM clouds has been a challenging endeavour: on the one hand, by looking at the H{\sc i} absorption against background quasars we can reach column densities well below $10^{17}$ cm$^{-2}$ but at the cost of having few single pencil-beam measurements per galaxy, given the sparse distribution of sufficiently bright background quasars \citep{tumlinson13,werk14,turner14,rubin15,lehner15,tumlinson17,tchernyshyov23,afruni23}; on the other hand, the detection of the H{\sc i} clouds in emission has always been limited by either spatial resolution or column density sensitivity.\\\indent Single-dish radio telescopes are able to reach limiting column densities of $10^{17}$ cm$^{-2}$ in H{\sc i} but with poor spatial resolution ($>200''$). Outside the Local Group, this size corresponds to $>3$ kpc and this is insufficient to constrain the nature of an isolated detection, i.e., it is not possible to distinguish between a primordial H{\sc i} cloud or a tiny dwarf galaxy, or a tidal feature \citep{fraternali01,thilker04,westmeier07,blok14,kerp16,veronese23}. Beam dilution can also wash out the signal from a gas cloud if it is very small with respect to the beam.\\\indent Previous generations of interferometric observatories achieved the necessary spatial resolution ($<100''$) but their column density sensitivity was usually not better than $\sim10^{19}$ cm$^{-2}$. Previous surveys (e.g., HIPASS, \citealt{hipass}; ALFALFA, \citealt{alfalfa}; THINGS, \citealt{things}; HALOGAS, \citealt{halogas}) have made significant contributions to the understanding of the H{\sc i} content in galaxies. However, for the reasons just described, they were limited in their ability to probe the diffuse atomic gas in the CGM and IGM, which remains largely unexplored.\\\indent The MeerKAT H{\sc i} Observations of Nearby Galactic Objects: Observing Southern Emitters (MHONGOOSE, \citealt{mhongoose2}) survey represents a recent advancement in observational capabilities, leveraging high sensitivity ($<10^{18}$ cm$^{-2}$) coupled with the excellent spatial resolution ($\sim30"$) and wide-field ($\sim1.5^\circ$) coverage of the MeerKAT radio telescope\footnote{The MeerKAT telescope is operated by the South African Radio Astronomy Observatory, which is a facility of the National Research Foundation, an agency of the Department of Science and Innovation.} \citep{meerkat}. The MHONGOOSE survey aims to detect and study the H{\sc i} gas in nearby star-forming galaxies, exploring their kinematics, gas dynamics, and environmental effects. The rich data set collected by MHONGOOSE also provides an opportunity to investigate the presence of faint signatures of pristine clouds embedded within the CGM/IGM. While this is usually done by attempting to find direct detections in data cubes with source finders, or even by eye,  we can also use alternative techniques that can push the sensitivity limit  for  the detection of faint HI structures around nearby galaxies even further down.\\\indent The most well-known of these is the spectral stacking, which consists of co-adding a large number of individual spectra with known redshifts to improve the Signal-to-Noise Ratio (SNR) of the final spectrum. H{\sc i} spectral stacking has been employed mainly to characterize the average H{\sc i} properties of samples of galaxies \citep{fabello11,delhaize13,gereb14,gereb15a,gereb15b,kanekar16,maccagni17,ianjam18,hiss,healy21c,chowdhury22,amiri23,apurba23}, but to date two studies used it to systematically look for H{\sc i} emission in the CGM/IGM around galaxies \citep{das20,das24}.\\\indent In this paper we will use this technique to search for H{\sc i} clouds around the MHONGOOSE galaxies but with a new approach. What differs from previous stacking experiments is that in this case we know the redshift of each galaxy, but there is no information about how the gas around them (i.e., outside the disk) is moving with respect to the disk, where it is located and what its morphology is. In other words, we will perform a kind of `unconstrained' stacking. For this reason, we use state-of-the-art cosmological simulations to find the best priors for our experiment, like the size of the H{\sc i} clouds and their velocity distribution, prior to applying the technique to the MHONGOOSE data. We also employ a one-dimensional spectral-line finder, based on the Source Finding Application (\texttt{SoFiA-2}, \citealt{sofia,sofia2}), designed to deal with low SNR lines at unknown spectral and spatial locations.\\\indent The rest of the paper is as follows. In Sect. \ref{sec:data} we describe the two datasets used in this paper: cosmological simulations that we use to test our methods and the observational data from the MHONGOOSE survey that we use for the actual stacking. The employed stacking and source finder algorithm is illustrated in Sect. \ref{sec:algo} and its testing and calibration with simulated data is provided in Sect. \ref{sec:testing}. We present its application to the MHONGOOSE galaxies and the results in Sect. \ref{sec:apply}, discussing also the statistical significance of the detected signals in the real data. A summary of our conclusions is provided in Sect. \ref{sec:conc}, while some future prospects are briefly outlined in Sect.\ref{sec:future}.

\begin{figure*}
    \resizebox{\hsize}{!}
    {\includegraphics[]{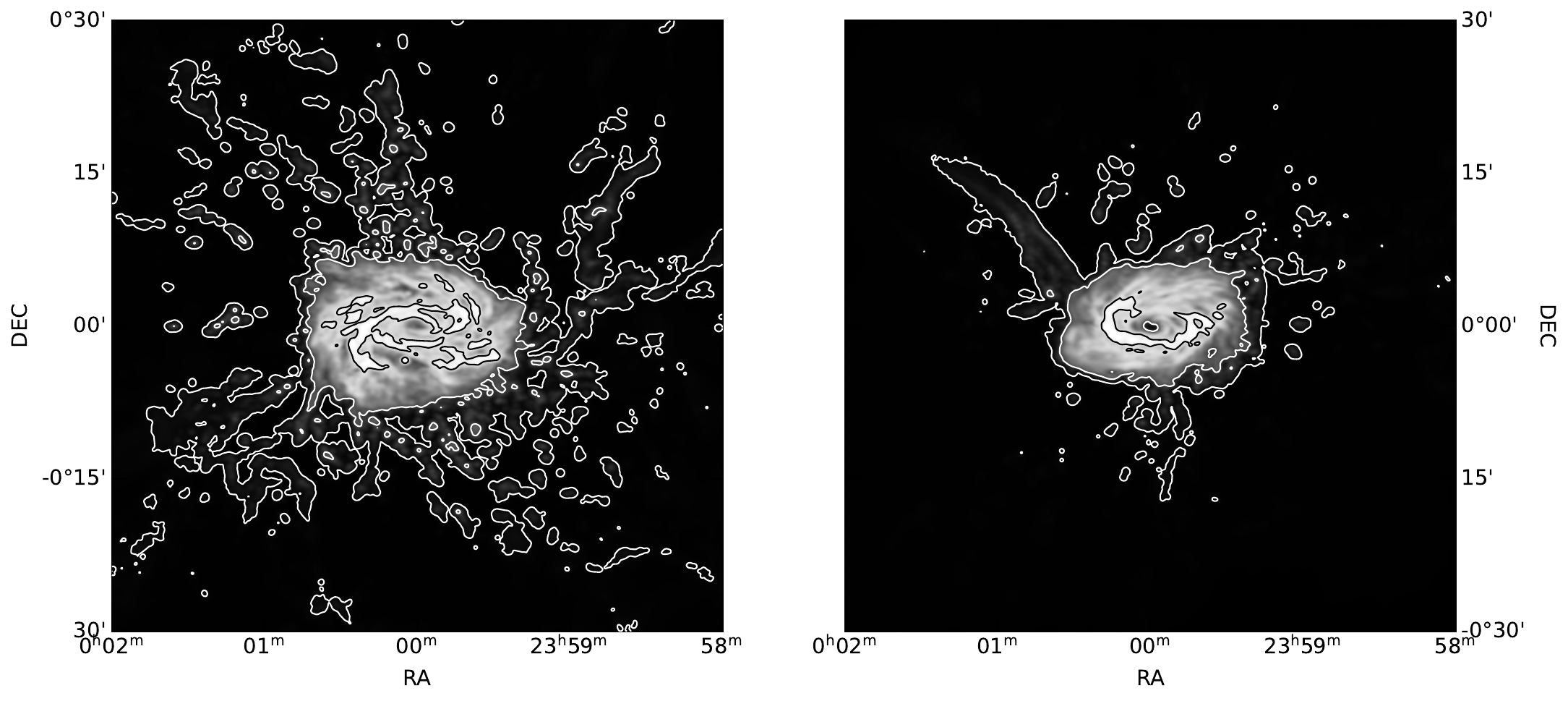}}
    \caption{Moment 0 map of the two TNG50 galaxies presented in this paper: 520855 on the left and 555013 on the right. Contours are denoting the $\text{Log}(N_{HI})=(16, 19, 21)$ cm$^{-2}$ column density in both panels. The field of view has a side-length of 350 kpc, sufficient to encompass the virial radius of both galaxies.}
    \label{fig:tng_mom0}
\end{figure*}

\begin{figure*}
    \resizebox{\hsize}{!}
    {\includegraphics[]{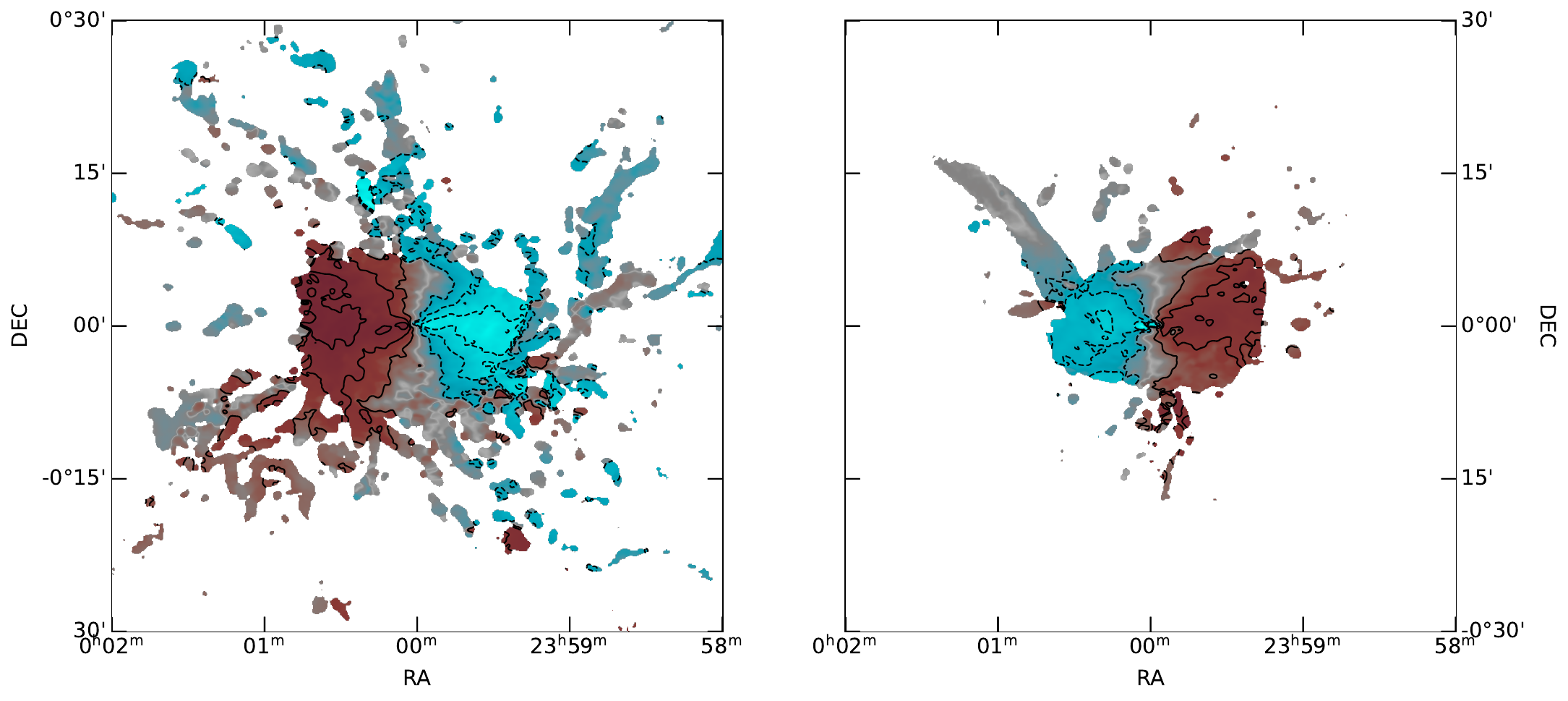}}
    \caption{Moment 1 map of the two TNG50 galaxies presented in this paper: 520855 on the left and 555013 on the right. Black solid-contours denote the receding (50, 100, 150) km s$^{-1}$ velocities with respect to the systemic velocity. Black dashed-contours instead refer to the approaching (-150, -100, -50) km s$^{-1}$ velocities. The maps are clipped at the $10^{16}$ cm$^{-2}$ column density level, as this is the noise level we expected to achieve via stacking.}
    \label{fig:tng_mom1}
\end{figure*}

\begin{table}[]
    \centering
    \caption{Main properties of the simulated galaxies used in this work.}
    \begin{tabular}{ c | c | c | c | c }
        \hline\hline
        ID & $M_{\text{halo}}$ & $M_*$ & $M_{\text{H{\sc i}}}$ & SFR \\
         & [$10^{12}$ M$_\odot$] &  [$10^{10}$ M$_\odot$] &  [$10^{10}$ M$_\odot$] & [M$_\odot$ yr$^{-1}$] \\
        \hline
        520885 & 1.17 & 5.47 & 3.17 & 2.76\\
        555013 & 0.89 & 3.66 & 1.85 & 1.48\\
        \hline
    \end{tabular}
    \tablefoot{The galaxies are part of the TNG50 Milky Way-like sample provided by \citet{ramesh23} and the H{\sc i} properties has been derived by Marasco et al. (in prep).\\The ID refers to the subhalo index in the group catalogues as determined by the subfind algorithm \citep{dolag09}. $M_{\text{halo}}$, $M_*$ and $M_{\text{H{\sc i}}}$ are the dark matter halo, stellar and H{\sc i} masses, respectively. SFR is the current star formation rate.}
    \label{table:tng}
\end{table}

\section{Dataset}\label{sec:data}
To test our algorithm we used mock H{\sc i} data based on the TNG50 simulation \citep{tng,ramesh23}, which predicts the presence of low-column density H{\sc i} clouds around galaxies. This enabled us to calibrate our algorithm and understand its limitations. Once the performance of our algorithm was assessed using these mock data, we applied it to the H{\sc i} cubes from the MHONGOOSE survey. In the following we briefly describe some of the properties of both the observed and simulated data.

\subsection{The MHONGOOSE sample}\label{sec:mhongoose}
The MHONGOOSE survey \citep{mhongoose2} using MeerKAT has performed ultra-deep H{\sc i} observations of 30 nearby gas-rich dwarf and spiral galaxies spanning a wide range of morphology, stellar mass ($4.7\leq\log M_*\leq10.7$ M$_\odot$) and star formation rate ($-2.52\leq\log\text{SFR}\leq0.75$ M$_\odot$ yr$^{-1}$). One of the main goals of the project is to identify and characterize the distribution of low column density H{\sc i}, which could potentially be inflowing towards the galaxies. These observations also seek to establish connections between this H{\sc i} and ongoing star formation processes within these galaxies.\\\indent Each of the 30 targets is observed for a total of 55 hours, spread across 10 observing sessions lasting 5.5 hours each. All ten observation tracks underwent the same calibration procedure and were combined to produce the `full-depth' cube. For every galaxy we also have a `single-track' cube, obtained from data reduction of a single 5.5 hour observation.\\\indent The calibration was performed using the Containerized Automated Radio Astronomy Calibration (CARAcal) pipeline \citep{caracal} as described in \citet{mhongoose2}. CARAcal provides a unified platform encompassing the standard calibration and reduction steps, including data flagging, cross-calibration, target separation, self-calibration, as well as continuum subtraction, spectral line imaging, and deconvolution.\\\indent The resultant H{\sc i} cubes were generated via WSClean \citep{wscleana,wscleanb} within the CARAcal pipeline. Cleaning of the H{\sc i} data followed a three-step iterative strategy which uses \texttt{SoFiA-2} to produce the cleaning masks. The process exploited the full MeerKAT resolution and sensitivity capabilities by producing six distinct H{\sc i} cubes for each galaxy with a 1.5$^\circ$ field of view (FoV). These cubes span a range of resolutions, from $7''$ to $90''$, achieved through variations in weighting and tapering.\\\indent As in this paper we aim to look for the H{\sc i} at the $\sim10^{17}$ cm$^{-2}$ column density level, we refrain using the high resolution cubes (\texttt{r00\_t00} and \texttt{r05\_t00} in \citealt{mhongoose2}) as this is not computationally efficient given their limiting column density being more than two order of magnitudes higher. The lowest resolution cubes (\texttt{r05\_t60} and \texttt{r10\_t90}) were also excluded given their sidelobe level being potentially able to affect the stacking. As the mock cubes described in Sect. \ref{sec:tng} are only used for testing purposes and a thorough comparison between TNG50 and MHONGOOSE will be provided in Marasco et al. (in prep.), between the intermediate resolution cubes \texttt{r10\_t00} and \texttt{r15\_t00} we decided for the latter because of its lower limiting column density ($3\sigma$, 16 km s$^{-1}$ sensitivity of $2.8\times10^{18}$ cm$^{-2}$). The beam size of $34.4''\times25.4''$ is slightly larger than for the mock cubes ($26''\times18''$), however, as described in Sect. \ref{sec:stacking} and Sect. \ref{sec:calibration} the effect of the beam will be mitigated via regridding and the stacking will be performed on scales much larger than the beam.\\\indent For the \texttt{r15\_t00} cubes we also used their derived \texttt{SoFiA-2} products obtained as described in \citet{mhongoose2}: zeroth moment (intensity), first moment (intensity-weighted line-of-sight velocity), and second moment (line-of-sight dispersion) maps, as well as the detection masks used to produce these maps.\\\indent For this work we used the single-track and full-depth cubes of the 18 MHONGOOSE galaxies having inclination angle $\leq60^\circ$. This selection is described in more detail in Sect. \ref{sec:assumption}.

\subsection{The TNG50 galaxies}\label{sec:tng}
The mock data used to test our algorithm were taken from IllustrisTNG \citep{tng}, a collection of extensive, cosmological simulations carried out using the \texttt{AREPO} moving-mesh code \citep{springel10}. The project comprises of three distinct simulation volumes (comoving box sizes: approximately 50, 100, and 300 Mpc on each side) and incorporates a comprehensive framework of galaxy formation physics, enabling a realistic representation of galaxy evolution across cosmic epochs \citep{weinberger17,pillepich18}.\\\indent For this work, we selected from the TNG50 Milky Way-like sample provided by \citet{ramesh23} two galaxies, 520855 and 555013\footnote{The ID refers to the subhalo index in the group catalogues as determined by the subfind algorithm \citep{dolag09}.} whose main properties are listed in Table \ref{table:tng}. In Fig. \ref{fig:tng_mom0} and Fig. \ref{fig:tng_mom1} we show their high-resolution H{\sc i} intensity and intensity-weighted line-of-sight velocity map\footnote{The velocity field colourmap used throughout this paper is taken from \citet{canvas}.}, respectively, and as observed with an inclination angle of 60 degrees. We chose these two galaxies because of their diverse H{\sc i} morphology, consisting of a well-defined H{\sc i} disk surrounded by clumpy and filamentary faint H{\sc i} features located up to $\sim100$ kpc of distance from the centre. These features are more evident for system 520855 (left panel of Fig. \ref{fig:tng_mom0}). The variety of H{\sc i} structures shown by these two galaxies make them optimal testing targets for the stacking algorithm we use in this paper.\\\indent The derivation of these H{\sc i} distributions is presented in Marasco et al. (in prep.). The goal of that paper is to derive the H{\sc i} distributions of a sample of simulated Milky Way-like galaxies, and to infer what they would have looked like if these systems were `observed' as a part of the MHONGOOSE survey. To achieve this, Marasco et al. (in prep.) produce synthetic observed H{\sc i} cubes for each simulated system matching the resolution, noise level and field of view of a real MHONGOOSE data set.\\\indent Here we use the synthetic data cubes for our two simulated test galaxies, derived with a robustness parameter of 1.0 (\texttt{r10\_t00}, cf \citet{mhongoose2}), giving a beam size of $26"\times18"$ and a $3\sigma$ column density sensitivity of $4.9\times10^{18}$ cm$^{-2}$ over 16 km s$^{-1}$. The simulated galaxies are all assumed to be at a distance of 20 Mpc, which is the average distance of the MHONGOOSE galaxies that Marasco et al. (in prep.) use as a comparison sample.

\begin{figure}
    \centering
    \includegraphics[width=\hsize]{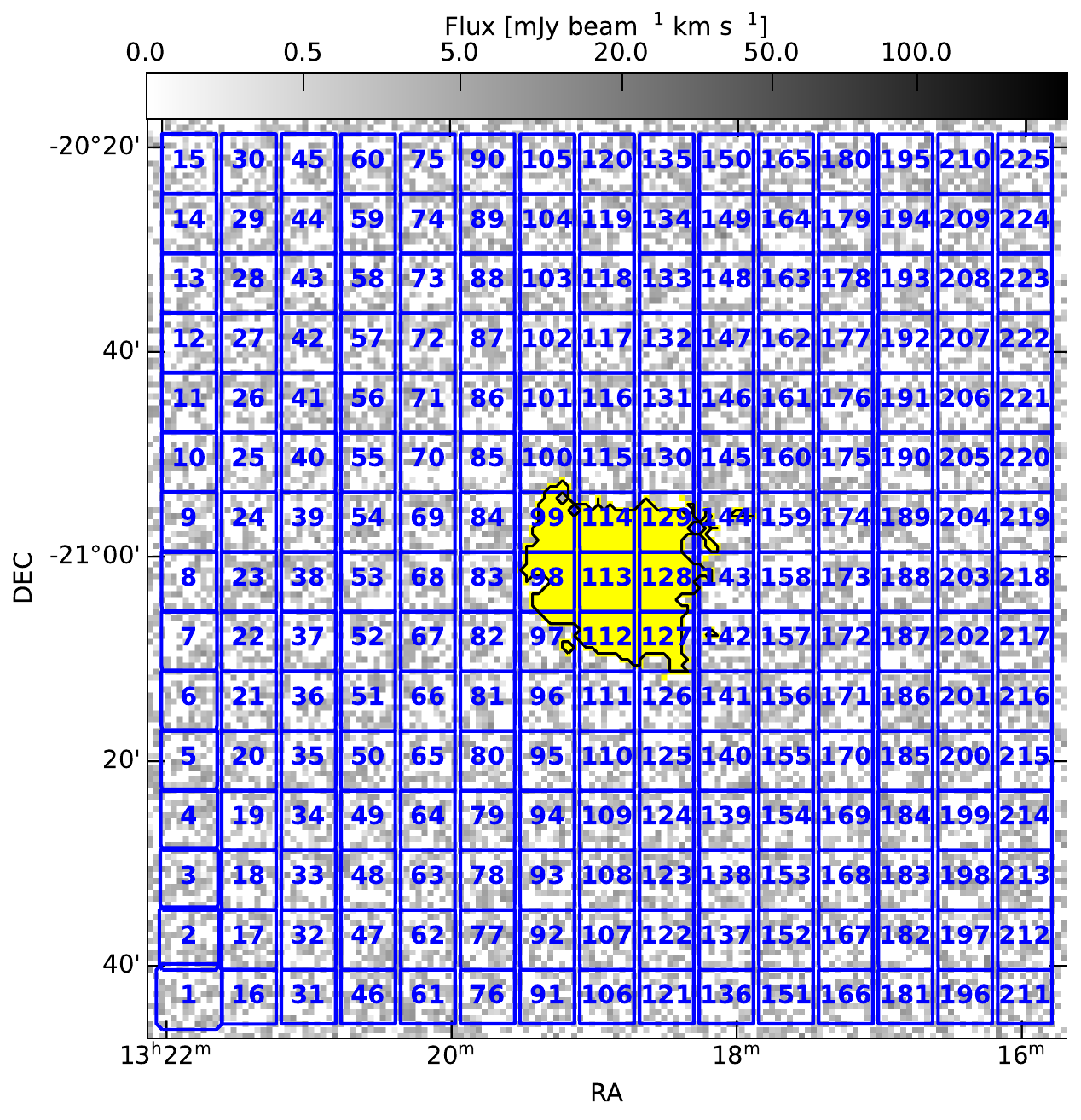}
    \caption{Stacking regions for J1318-21 reported as blue squares. For the cell size displayed in the figure, \texttt{STACKER} will provide 225 stacked spectra, one for each region. The numbering helps for the bookkeeping. The background grey-scale image is the cube resulting from the preliminary manipulation steps (i.e., masking, shuffling and regridding) collapsed along the spectral axis and the black contours enclose the \texttt{SoFiA-2} mask denote with a yellow colour.}
    \label{fig:regions}
\end{figure}

\begin{figure*}
    \resizebox{\hsize}{!}
    {\includegraphics[]{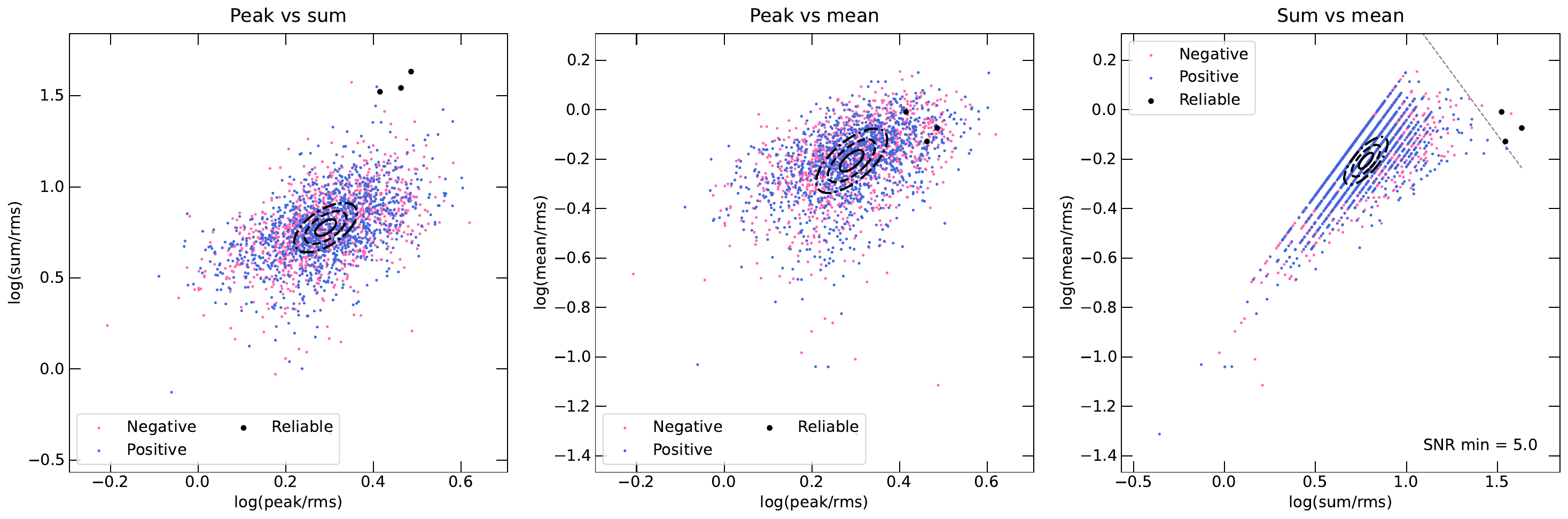}}
    \caption{Reliability plot of the sources detected in J1318-21 full-depth cube (intermediate cell size). \textit{Left panel}: $\frac{F_{sum}}{\sigma}$ vs. $\frac{F_{peak}}{\sigma}$ projection of the ($\frac{F_{max}}{\sigma}$, $\frac{F_{sum}}{\sigma}$ and $\frac{F_{mean}}{\sigma}$) parameter space. Blue, pink and black points are the positive, negative and reliable sources, respectively, detected in the 225 stacked spectra computed for the J1318-21 full-depth cube when using the intermediate cell size. The solid, dashed and dash-dotted black ellipses represent, respectively, the two-dimensional 1, 2 and 3 times the standard deviation of the source distribution. \textit{Centre panel}: $\frac{F_{mean}}{\sigma}$ vs. $\frac{F_{peak}}{\sigma}$ projection. \textit{Right panel}: $\frac{F_{mean}}{\sigma}$ vs. $\frac{F_{sum}}{\sigma}$ projection. The dashed dark-grey line is the minimum SNR of a source in order to be considered reliable.}
    \label{fig:reliability}
\end{figure*}

\section{The algorithm}\label{sec:algo}
The algorithm we use here to search for H{\sc i} signatures consists of a spectral stacking routine combined with a one-dimensional spectral-line finder. These two functions are wrapped into a publicly-available \texttt{Python} package\footnote{https://github.com/SJVeronese/nicci-package.}. Throughout the paper we refer to the spectral stacking task as \texttt{STACKER} and to the line finding task as \texttt{FINDER}. In the following sections we describe each task in detail and we refer to Sect. \ref{sec:calibration} for details on how the values of the input parameters were determined.

\subsection{Spectral stacking}\label{sec:stacking}
\texttt{STACKER} begins with several preliminary manipulations on the input data cube:
\begin{itemize}
    \item it blanks specific lines-of-sight (LoS) based on a supplied mask to avoid stacking the emission from the main galaxy and other already-known sources. For this we used the \texttt{SoFiA-2} detection mask that was also used to create the moment maps as described in \citet{mhongoose2};
    \item it aligns the spectra according to a velocity field, a process called `shuffle'. It consists of shifting each spaxel along the spectral axis so that its central channel corresponds to the value of the same LoS in the velocity field. To preserve the noise properties of the data, the shifting wraps the spectrum around\footnote{The spectral alignment is performed using our implementation of the \texttt{GIPSY} \citep{gipsy,gipsy2} task \texttt{SHUFFLE}.}, meaning it inserts the low frequencies channels at the high frequency end. For the TNG50 galaxies we initially used the noiseless moment 1 map, as it contains the kinematical information of the gas outside the \texttt{SoFiA-2} detection mask (see Sect. \ref{sec:testing}). As this information is missing when dealing with real data, we tested several velocity field assumptions as described in Sect. \ref{sec:assumption};
    \item it regrids the resulting shuffled cube to mitigate the spatial correlation of the voxels, so that the noise decreases as the square root of the number of co-added spectra (under the assumption of Gaussian noise). To achieve this, the pixel size is increased to match the beam size\footnote{The regridding is done via the \texttt{reproject\_interp} function from the \texttt{Astropy Reproject} package \citep{astropy13,astropy18,astropy22}. The interpolation is done using a nearest-neighbour method to preserve the underlying noise properties. Using different interpolation orders or \texttt{Reproject} functions leads to negligible improvement, but a significant cost in terms of computing time.}. 
\end{itemize}
We do not apply any primary beam correction at this stage because stacking requires the noise to be as uniform as possible. A primary beam correction increases the $\sigma$ level toward the outskirts of the FoV.\\\indent After these preliminary manipulations, \texttt{STACKER} assigns each position on the sky to a stacking region, that we choose to be square cells (see Fig. \ref{fig:regions}). The stacked spectrum for each region is then computed as follow:
\begin{equation}\label{eq:stack}
    F_S=\frac{\sum_iF_i\cdot w_i}{\sum_i w_i}
,\end{equation}
where $F_S$ is the stacked flux in a given channel, calculated by summing the fluxes ($F_i$) of the spectra in that channel, weighted by their respective weights ($w_i$). The common weighting scheme used in the literature is the so-called `$\sigma^2$-weight' (e.g., \citealt{fabello11,kanekar16,ianjam18,hiss,healy21c,chowdhury22,apurba23}): $w_i=\frac{1}{\sigma_i^2}$, where $\sigma_i$ represents the $i^{th}$-spectrum's noise level. As the presence of signal hidden in the noise potentially increases $\sigma_i$, leading to the down weighting of the corresponding spectrum and thus reducing the contribution of any hidden signal, we also considered the `equal-weight' scheme ($w_i=1$ $\forall$ $i$) and tested against the $\sigma^2$-weight as described in Sect. \ref{sec:calibration}.\\\indent The stacked spectra, one for each region, are passed to \texttt{FINDER} for source finding.

\begin{table*}[]
    \centering
    \caption{Source finder parameters used on the TNG50 cubes.}
    \begin{tabular}{l  c | c | c | c }
        \hline\hline
        \multicolumn{2}{ c }{Parameters} & \multicolumn{3}{ c }{Cell size} \\ 
        && $14\times14$ beams & $9\times9$ beams & $5\times5$ beams \\ 
        \hline
        Flux threshold & [$\sigma$] & 2.5 & 2.5 & 2.5 \\ 
        Smoothing kernels & [channels] & 1, 5, 10 & 1, 5, 11 & 1, 7, 12 \\ 
        Minimum linewidth & [channels] & 5 & 5 & 7 \\ 
        Minimum source SNR && 5 & 5 & 5 \\ 
        Reliability threshold && 0.85 & 0.85 & 0.85 \\ 
        \hline
    \end{tabular}
    \label{table:tngfinder}
\end{table*}

\subsection{Source finding}\label{sec:finder}
\texttt{FINDER} employs a simplified one-dimensional version of the \texttt{SOFIA-2} smooth-and-clip (S$+$C) algorithm, a procedure involving smoothing the data on multiple scales and apply for each scale a $\sigma$-clip \citep{serra12b}. The output is binary masks, one for each stacked spectrum, which are subsequently used to create the source catalogue by combining the detected channels in each mask, i.e., assigning consecutive detections to a single source. Positive and negative sources spanning fewer than a given number of channels are removed from the catalogue.\\\indent Following the method described in detail by \citet{serra12}, \texttt{FINDER} computes the reliability of the remaining sources to discriminate genuine lines from noise peaks. Each source is placed in a three-dimensional space ($\frac{F_{max}}{\sigma}$, $\frac{F_{sum}}{\sigma}$ and $\frac{F_{mean}}{\sigma}$), where $F_{max}$, $F_{sum}$ and $F_{mean}$ are the peak, total and mean flux (see Fig. \ref{fig:reliability} for an example of the parameter space plot). 
For each positive source $P_i=(F_{i,max},F_{i,sum},F_{i,mean})$ a volume density of positive ($n_{pos}$) and negative ($n_{neg}$) sources, as determined on the collection of stacked spectra, is calculated through the Gaussian kernel density estimation \citep{hastie01}:
\begin{equation}
    n=\frac{1}{h}\sum_i\mathcal{G}\left(\frac{x-x_i}{h}\right)
,\end{equation}
where $x=(x_1,x_2,...,x_n)$ are the position of the sources, $\mathcal{G}$ is a Gaussian kernel and $h$ is a smoothing factor. The reliability ($\varrho$) of $P_i$ is then given by
\begin{equation}
        \varrho=\begin{cases}
        \frac{n_{pos}-n_{neg}}{n_{pos}}, &\text{for }n_{pos}\geq n_{neg}\\ 
        0, &\text{otherwise}\\ 
    \end{cases}
.\end{equation}
Only sources for which $\varrho$ is above a given threshold are retained in the final source catalogue.

\subsection{Gaussianity tests}\label{sec:tests}
MeerKAT noise is known to be Gaussian to a high degree, as illustrated in Sect. \ref{sec:noise}. To quantify this further we incorporated two Gaussianity tests in \texttt{FINDER}. The first is the Anderson-Darling test \citep{stephens74,stephens76,stephens77,stephens79}, which evaluates the hypothesis that a given sample follows a specific distribution, in our case, a Gaussian distribution. This statistical test quantifies the disparity between the observed data and the expected values by computing the Anderson-Darling statistic $A$
\begin{equation}
    A^2=-n-\sum _{i=1}^{n}{\frac {2i-1}{n}}[\ln(F(x_{i}))+\ln(1-F(x_{n+1-i}))]
,\end{equation}
where $n$ is the number of elements in the sample, $F$ is the hypothesized cumulative distribution function (in our case, a Gaussian) and $x_i$ is the $i^{th}$-datapoint in progressive order ($x_1<\cdots<x_{n}$).\\\indent A lower Anderson-Darling statistic indicates a better fit to the Gaussian distribution, suggesting that the data can reasonably be assumed to follow it. Conversely, a higher statistic indicates a poorer fit, meaning non-Gaussian behaviour. A p-value
\begin{equation}
    p=\begin{cases}
        e^{1.2937-5.709A+0.0186A^2}, &\text{if }A\geq0.6\\ 
        e^{0.9177-4.279A-1.38A^2}, &\text{if }0.34\leq A<0.6\\ 
        e^{-8.318+42.796A-59.938A^2}, &\text{if }0.2\leq A<0.34\\ 
        e^{13.436+101.14A-223.73A^2}, &\text{otherwise}\\ 
    \end{cases}
,\end{equation}
is associated with the outcome of the test. Since $p=0.05$ is a `standard' value for rejecting the alternative hypothesis, in our case spectra for which $p<0.05$ are considered to have non-Gaussian noise.\\\indent We complement the Anderson-Darling with another Gaussianity test which involves smoothing the spectrum using a boxcar filter with a width of 10 channels (N. Kanekar, private comm.). The $\sigma$ before and after the smoothing is then compared, with the expectation that it should decrease by a factor of $\sqrt{10}$ if the noise adheres to a Gaussian distribution. Detections from spectra that do not satisfy either of these conditions are rejected from the final source catalogue.

\begin{figure*}
    \resizebox{\hsize}{!}
    {\includegraphics[]{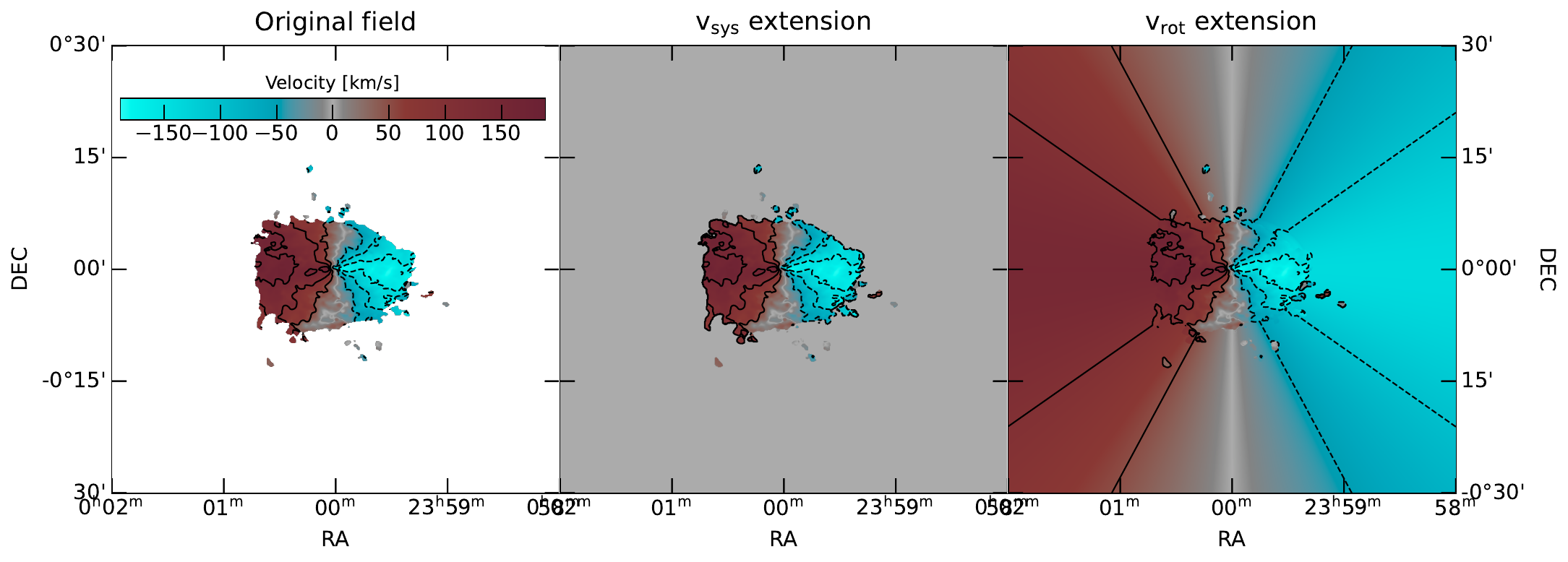}}
    \caption{Comparison between differently extended velocity fields. \textit{Left panel}: \texttt{SoFiA-2} moment 1 map of the TNG50 galaxy 520855. Black solid-contours denote receding (50, 100, 150) km s$^{-1}$ velocities with respect to the systemic velocity. Black dashed-contours instead refer to approaching (-150, -100, -50) km s$^{-1}$ velocities. \textit{Centre panel}: Velocity field extended with the systemic velocity. \textit{Right panel}: Velocity field extended with the rotation velocity of the H{\sc i} disk.}
    \label{fig:velfi}
\end{figure*}

\begin{figure}
    \centering
    \includegraphics[width=\hsize]{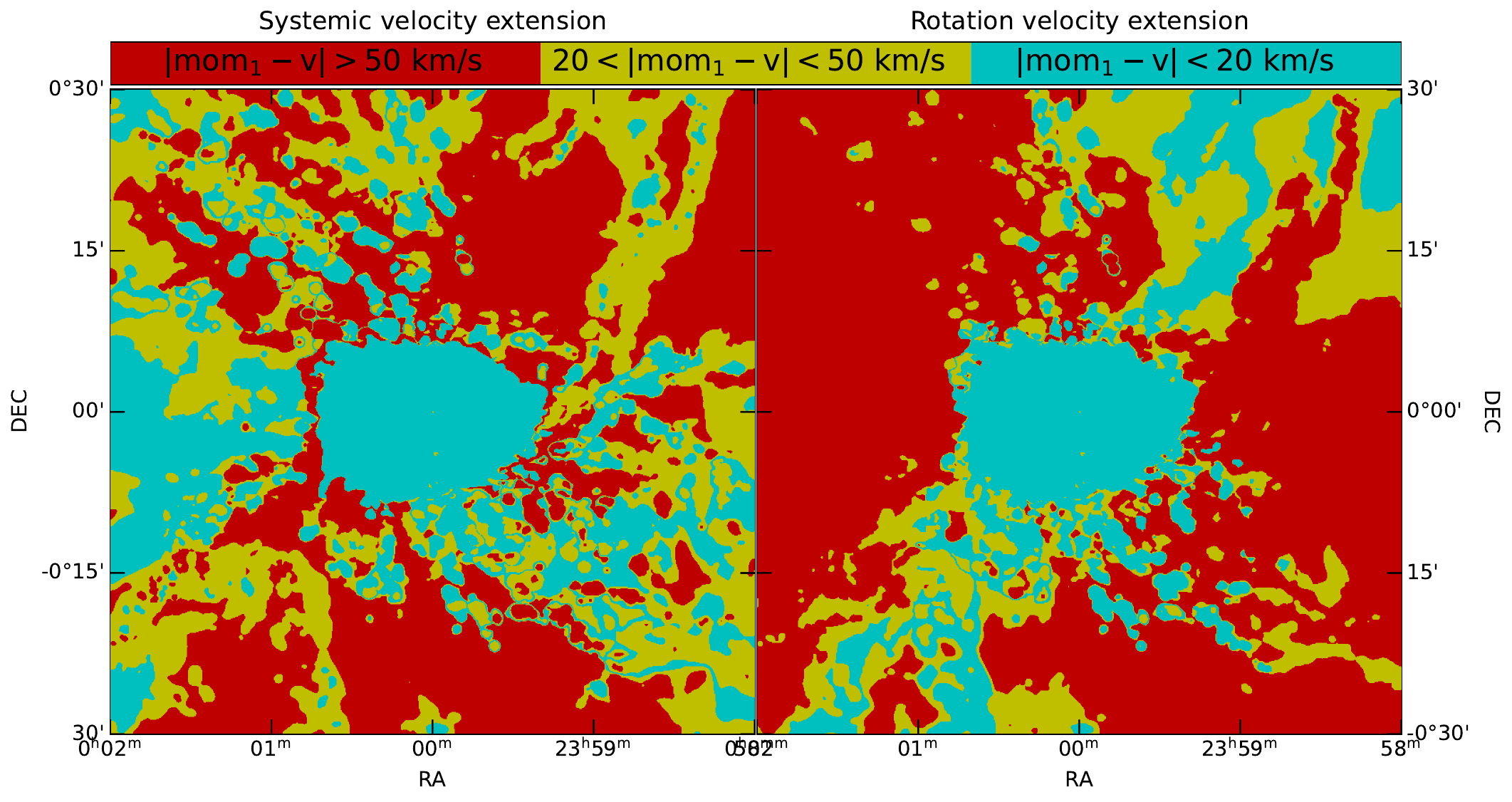}
    \caption{Error in the spectral alignment for the TNG50 galaxy 520885 under two different CGM/IGM kinematics assumptions. \textit{Left panel}: the residuals in terms of $|\text{mom}_1-v|$, where $\text{mom}_1$ is the noiseless moment 1 map and $v$ the assumed kinematics, when $v=v_{sys}$. Light-blue corresponds to $|\text{mom}_1-v|<20$ km s$^{-1}$, yellow to $20<|\text{mom}_1-v|<50$ km s$^{-1}$ and red is $|\text{mom}_1-v|>50$ km s$^{-1}$. \textit{Right panel}: same as left panel but when $v$ is given by assuming the gas is co-rotating with the galaxy with a flat rotation curve.}
    \label{fig:alignment1}
\end{figure}

\section{Testing the method}\label{sec:testing}
Both \texttt{STACKER} and \texttt{FINDER} depend on a number of parameters which need to be optimized, mainly, the weighting scheme, the size of the square cells, the smooth-and-clip scales and the reliability threshold. We tested and optimised our procedure using simulated galaxies, which provide precise information about the priors, i.e., location, column density, and kinematics of the gas. Therefore, we performed the stacking on the two TNG50 cubes described in Sect. \ref{sec:tng}.\\\indent The emission from the galaxy was blanked based on the \texttt{SoFiA-2} mask and the cubes were shuffled using the noiseless moment 1 map, which also contains kinematic information on the low column density gas otherwise hidden in the noise. The latter choice is to maximise the SNR of each stacked spectrum, as we are assuming to know precisely the kinematics of the gas. In practice, these velocities are not precisely known, thus we test the effect of these uncertainties in Sect. \ref{sec:assumption}.\\\indent We then regrid the shuffled cube to the beam FWHM. As already mentioned in Sect. \ref{sec:stacking}, this will make voxels (almost) independent and will allow the noise to decrease with the square root of the number of co-added LoS.

\subsection{Parameters optimization}\label{sec:calibration}
We first optimised the size of the stacking regions, aiming to balance the number of stacked spectra with the filling factor of a source. The noise ($\sigma_s$) in the stacked spectrum decreases as $\sqrt{N}$, where $N$ is the number of co-added LoS. Increasing the cell size, i.e., increasing $N$ leads on the one hand to a lower $\sigma_s$, but on the other hand it might potentially lead to the inclusion of more noise-only spectra which wash out the signal. Conversely, reducing the cell size, i.e., decreasing $N$, will increase $\sigma_s$.\\\indent Given that the noise decreases as $\sqrt{N}$, we set the smallest cell size to be $5\times5$ beams, which corresponds to an improvement in the limiting column density of up to a factor of 5. It is not worth opting for smaller cell sizes because of the otherwise negligible improvement in sensitivity. We then progressively increased the cell size until the number of detections is maximised. This results in choosing the largest size to be $14\times14$ beams, equivalent to an improvement in the limiting column density by a factor of 14. We also employed an intermediate size of $9\times9$ beams.\\\indent In a general cosmology with $H_0=69.6$ km s$^{-1}$ Mpc$^{-1}$, $\Omega_m=0.286$ and $\Omega_v=0.714$ \citep{bennet14}, for a galaxy at 22.7 Mpc (the furthest in the MHONGOOSE sample) observed with the \texttt{r15\_t00} beam (see Sect. \ref{sec:mhongoose}) the cell sizes correspond to physical scales of $52\times52$, $34\times34$ and $19\times19$ kpc$^{2}$, respectively, while for a galaxy at 3.3 Mpc (the closest in the MHONGOOSE sample) to $8\times8$, $5\times5$ and $3\times3$ kpc$^{2}$, respectively. These scales overlap with the predicted physical size of H{\sc i} clouds in the CGM \citep{ramesh24}. Given the wide range in distances, keeping the cell size fixed to a set of physical scales is not feasible for our stacking experiment, thus, we fixed them instead to the aforementioned values of $14\times14$, $9\times9$ and $5\times5$ beams.\\\indent We choose the weighting scheme (see Sect. \ref{sec:stacking}) by comparing the outcomes when employing the equal- and the $\sigma^2$-weight. The difference was negligible, both in terms of $\sigma_s$ and source detection, so we ended up choosing the equal-weight for the reasons provided at the end of Sect. \ref{sec:stacking}. Because the noise injected in the mock cubes is uniform by construction (see Sect. \ref{sec:tng}), it is not surprising that the two weighting schemes lead to comparable results.\\\indent Lastly, we optimised the \texttt{FINDER} specifications, namely, the $\sigma$-threshold, the smoothing kernels, the linker size and the reliability parameters. We found that imposing $\sigma\text{-threshold}=2.5$ allows for the detection of a sufficient number of noise peaks to provide a robust reliability calculation.\\\indent The optimal smoothing kernels, linker size and reliability parameters were selected by changing each of them until the difference between the observed and expected cumulative distribution of the Skellam variable
\begin{equation}\label{eq:skellam}
    K=\frac{n_{pos}-n_{neg}}{\sqrt{n_{pos}+n_{neg}}}
\end{equation}
is minimized. A thorough description of $K$ is provided by \citet{serra12}.\\\indent In Table \ref{table:tngfinder} we list the values employed in \texttt{FINDER} to detect signal in the stacked spectra around the two simulated galaxies.

\subsection{Assumption on the gas kinematics}\label{sec:assumption}
So far we have aligned the spectra with the noiseless moment 1 map, i.e., assuming to know precisely the kinematics of the gas. As this will not be the case when stacking the MHONGOOSE data, we proceeded by evaluating two possible assumptions about the gas kinematics that are used as a prior: systemic velocity, which is equivalent to no alignment, or flat rotation curve. The former represents the scenario where the gas outside the galaxy is assumed to be unaffected by the rotation of the galaxy, the latter assumes that the CGM/IGM co-rotates with the H{\sc i} disk.\\\indent Ideally, to test the second hypothesis one wants the rotation curve to be derived and correctly extrapolated to large radii, i.e., beyond the `edge' of the disk as identified in the moment maps. This prior, namely the geometry of the gaseous disk as a function of radius, is generally unknown. Exploring the outcome of stacking when the spectral alignment is done with the geometry and the rotation velocity as a free parameter is computationally expensive and beyond the scope of this paper. It would, however, be a natural follow-up of this work (see Sect. \ref{sec:future}).\\\indent To evaluate the two hypothesis we extrapolated to larger radii the moment 1 map of the two TNG50 galaxies as produced by \texttt{SoFiA-2}. For the case where the gas is assumed to move at the systemic velocity, this is done by simply replacing any empty value in the map with the systemic velocity of the galaxy. For the assumption that the gas is co-rotating with the galaxy, the extension of the moment 1 map is less trivial. We retrieved the values of the moment 1 map along the major axis ($v_{rot}$) and, together with the azimuthal angle ($\theta$) and inclination ($i$) of the galaxy, we computed the velocity field across the entire FoV:
\begin{equation}\label{eq:v}
    v=v_{sys}+v_{rot}\cos{\theta}\sin{i}
,\end{equation}
for which we provide in Fig. \ref{fig:velfi} the result for the galaxy 520885\footnote{The extension of the velocity field is performed using our implementation of the \texttt{GIPSY} task \texttt{VELFI}.}.\\\indent We then calculated for each galaxy the residual maps $R=|\text{mom}_1-v|$, where $\text{mom}_1$ is the noiseless moment 1 and $v$ is the extended velocity field under one of the two aforementioned assumptions. The residual maps for the galaxy 520885 are provided in Fig. \ref{fig:alignment1}, while the residuals for the galaxy 555013 are given in Appendix \ref{sec:app1}. Since $R$ corresponds to the alignment error for each LoS, by looking at Fig. \ref{fig:alignment1} one can see where each assumption will produce too large an error ($>50$ km s$^{-1}$, \citealt{khandai11,maddox13,neumann23}) in the spectral alignment. In this case using the systemic velocity generally provides a better result.\\\indent It will be clear that one should be careful not to generalise this result. However, we can use the inclination value to our advantage when applying the stacking. If $i$ is the inclination angle and $v_{rot}$ the rotation velocity of the gas, the maximum velocities measured along each LoS are related approximately by Eq. (\ref{eq:v}). For a face-on galaxy ($i=0^\circ$), $v\sim v_{sys}$ no matter the rotation velocity. If the gas accretion happens along the plane of the disk, assuming the H{\sc i} clouds are moving at the systemic velocity is a reasonable choice. For an edge-on galaxy, $v_{rot}$ is the dominant factor in Eq. (\ref{eq:v}), hence, any uncertainty in the estimate of the kinematics is likely to produce a larger error in the alignment. Again, a study of the most suitable assumption on the gas kinematics as a function of the geometrical properties of the galaxy is beyond the scope of this work and will be analysed in a forthcoming paper.\\\indent In this paper we employ the systemic velocity as the assumed kinematics of the gas in the CGM/IGM for stacking the MHONGOOSE data and we apply our algorithm to the 18 galaxies with $i\leq60^\circ$.\\\indent A last point to consider for the quality of the alignment is the coherency of the assumed kinematics with respect to the real one on the scales probed by the chosen cell sizes. For instance, if in a given cell size the assumed kinematics cover only a small velocity range, then even if the spectra in the cell are assigned erroneous velocity shifts, this will still result in a coherently aligned stacked profile, just with an extra velocity shift with respect to the reference velocity. However, measuring this coherency in the absence of the prior kinematics, as when we stack the MHONGOOSE cubes, is not possible.

\begin{figure}
    \centering
    \includegraphics[width=\hsize]{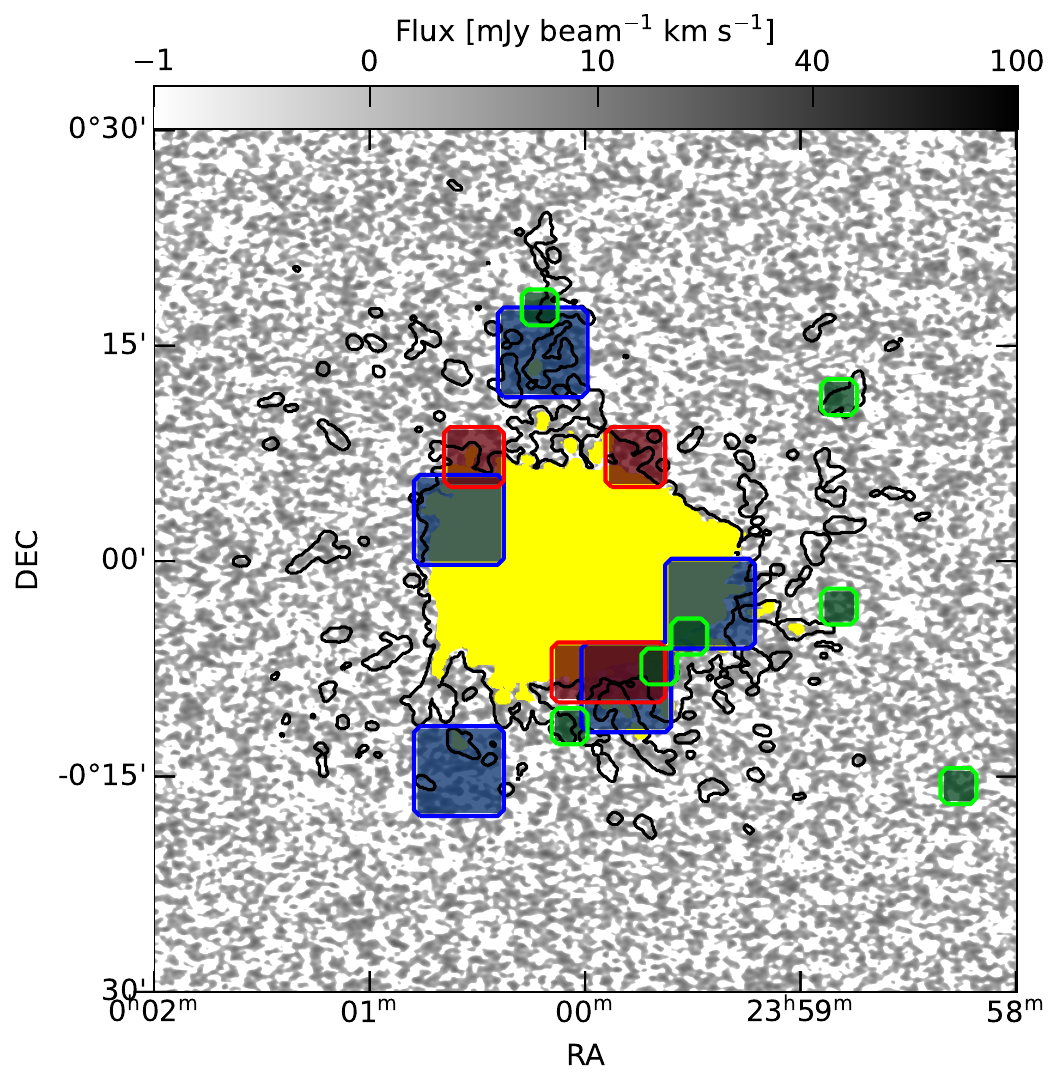}
    \caption{Stacking detection map. The background grey-scale image is the cube of the TNG50 520855 galaxy collapsed along the spectral axis and blanked from the galaxy emission. Black contours denote the noiseless moment 0 map clipped at the column density value of $3.6\times10^{17}$ cm$^{-2}$. The coloured squares indicate the stacking regions where their stacked spectrum contains a reliable detection. Different colours represent different cell size: blue for $14\times14$, red for $9\times9$ and green for $5\times5$ beams.}
    \label{fig:cal}
\end{figure}

\begin{figure*}
    \resizebox{\hsize}{!}
    {\includegraphics[]{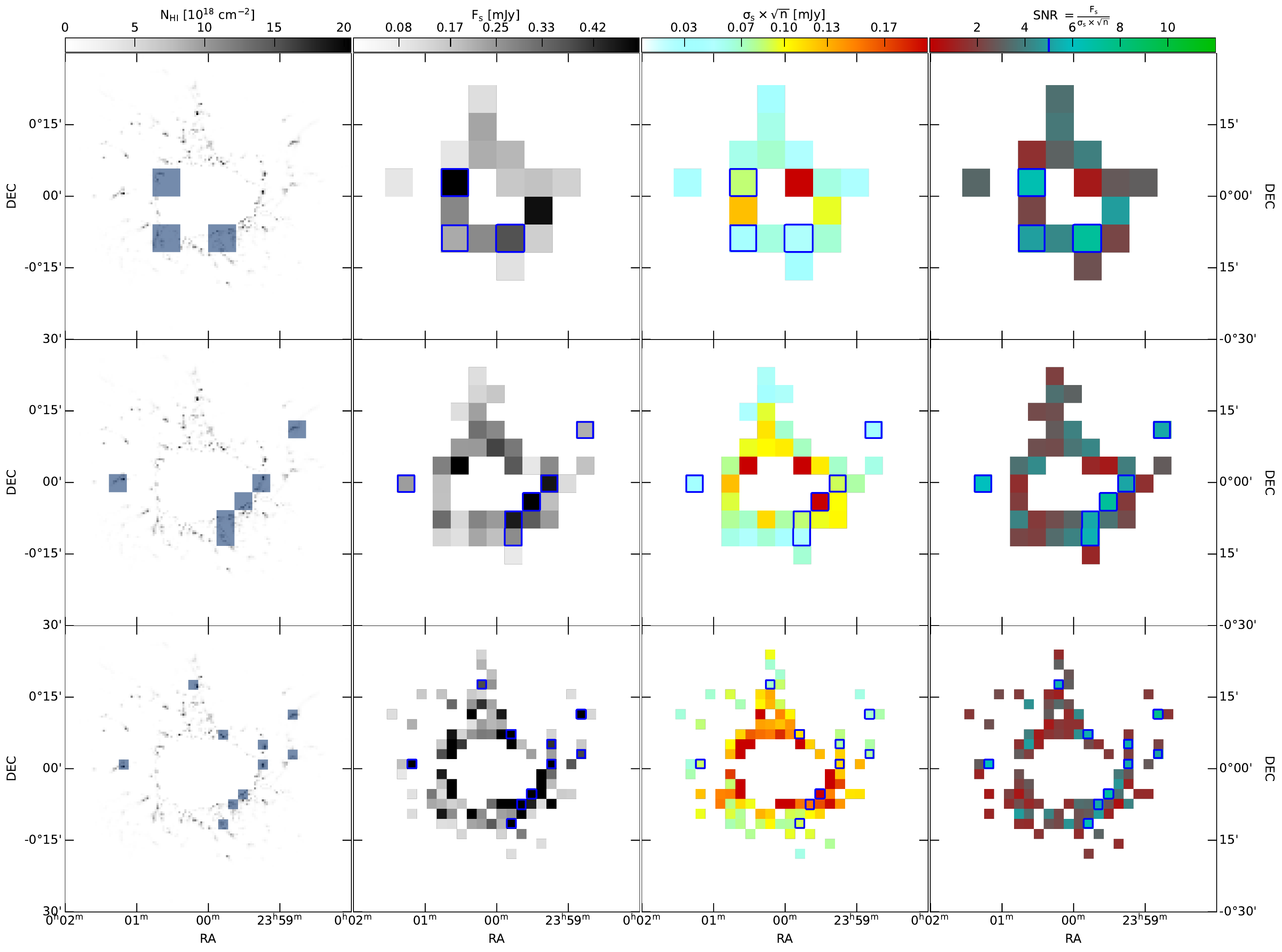}}
    \caption{Stacking SNR maps for the TNG50 galaxy 520885. \textit{First column}: stacking regions where SNR $>5$ (blue squares) overlaid with the grey-scale noiseless moment 0 map, blanked from the emission already detected by \texttt{SoFiA-2}. Top to bottom is for cell sizes of $14\times14$, $9\times9$ and $5\times5$ beams, respectively. \textit{Second column}: the integrated flux of the signal for stacking regions where the H{\sc i} column density is $>3.6\times10^{17}$ cm$^{-2}$. The blue contours enclose the cells where SNR $>5$. \textit{Third column}: the noise of the stacked spectrum multiplied by the square root of the number of channels covered by the source. \textit{Fourth column}: integrated SNR of the signal in the stacked spectrum for the regions. The SNR $=5$ level is indicated also on the colourbar.}
    \label{fig:snrmaps}
\end{figure*}

\subsection{The limitation of our method: the effect of the SNR cut}\label{sec:limit}
Fig. \ref{fig:cal} shows the results of the stacking experiment for TNG50 520855 galaxy where the alignment has been done using the systemic velocity. Regions where a reliable detection has been found are denoted with the blue, red and green boxes, corresponding to cell sizes of $14\times14$, $9\times9$ and $5\times5$ beams, respectively. The boxes are overlaid with the background grey-scale image resulting from the collapse of the cube along the spectral axis. The galaxy emission is blanked according to the \texttt{SoFiA-2} mask marked in yellow. Since this emission was blanked prior to the stacking, the sources in the coloured boxes represent signal below the \texttt{SoFiA-2} detection threshold of $4\sigma$.\\\indent The 2.5$\sigma$-16 km s$^{-1}$ column density limit that we were able to achieve with stacking is $1.3\times10^{17}$ cm$^{-2}$ for the largest cell size, $2.1\times10^{17}$ cm$^{-2}$ for the intermediate size and $3.6\times10^{17}$ cm$^{-2}$ for the smallest size. However, only $\sim35\%$ of the emission above the $3.6\times10^{17}$ cm$^{-2}$ column density level, shown as black contours in Fig. \ref{fig:cal}, was detected.\\\indent The reason why not all emission was detected is that stacking gives a weighted mean, hence, a line intrinsically bright enough to potentially fall above the detection threshold, will become too faint to be picked up if averaged with only noise. Furthermore, through the reliability procedure we introduce a filtering based on the SNR of a source: detections with SNR lower than a threshold are automatically rejected and do not enter the detection map (see the grey dashed line of the top-right panel in Fig. \ref{fig:reliability}).\\\indent Therefore, we tested, given our set of filtering parameters, what our stacking algorithm should be able to pick up. We impose an integrated SNR cut of 5 (see Table \ref{table:tngfinder}), that with our spectral resolution of 1.4 km s$^{-1}$ (see Sect. \ref{sec:tng}) corresponds to a per-channel SNR of 1.5 for a linewidth of 16 km s$^{-1}$.\\\indent We proceeded to quantify the amount of gas that will fall below our chosen SNR threshold, as illustrated in Fig. \ref{fig:snrmaps}. The integrated SNR for a source spanning $n$ channels in a stacked spectrum having noise $\sigma_s$ is defined as:
\begin{equation}\label{eq:snr}
    \text{SNR}=\frac{\sum F_i}{\sigma_s\sqrt{n}}
,\end{equation}
where $F_i$ is the flux in the $i$-th channel covered by the source. We computed equation (\ref{eq:snr}) for each stacking region by running \texttt{STACKER} on the noiseless cube. We retrieved first the dispersion of the line in every stacked spectrum, which gives us $n$, and then by integrating each noiseless stacked spectrum within a window of size $n$ around its peak we calculated $\sum F_i$. Finally, $\sigma_s$ is retrieved from the `noisy' stacked spectrum. The result is a map with the SNR for each stacking region.\\\indent Clipping this map at $\text{SNR}>5$ provides the regions where the signal in their stacked spectrum is bright enough to be detected given the SNR cut we have imposed. This is illustrated in Fig. \ref{fig:snrmaps}, where we show the outcome for the TNG50 galaxy 520885 (the result is analogous for the other mock galaxy as shown in Appendix \ref{sec:app1}). The blue squares are the regions where $\text{SNR}>5$.\\\indent Consequently, the reason why in Fig. \ref{fig:cal} not all the emission above the $3.6\times10^{17}$ cm$^{-2}$ level (black contours) has been detected with stacking is the following: if the mean flux within each region is not enough to survive the imposed SNR cut, then that flux will not be detected, no matter how it is distributed within the region.

\begin{figure}
    \centering
    \includegraphics[width=\hsize]{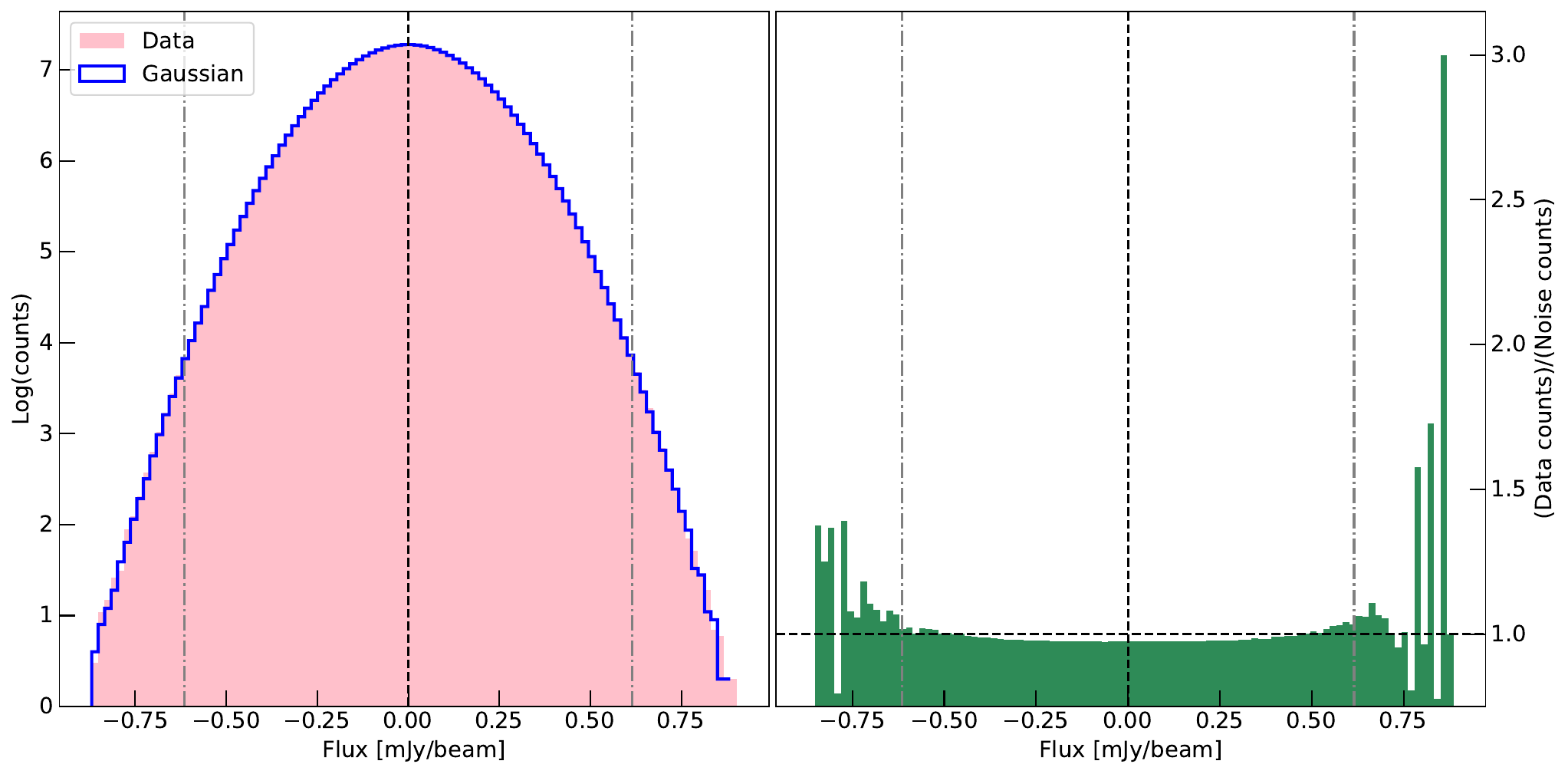}
    \caption{Comparison between the MeerKAT and Gaussian noise distribution. \textit{Left panel}: the pink linear-log histogram represent the number of voxels with a given flux value in a MeerKAT cube, known to contain no bright galaxies. The blue linear-log histogram is instead computed from a Gaussian noise cube. The black dashed vertical line is the reference 0-flux level, whereas the two grey dash-dotted lines are denoting the $\pm4\sigma$ level. \textit{Right panel}: ratio between the pink and blue histograms of the left panel. If the MeerKAT noise is truly Gaussian, the ratio should be $\sim1$ for every flux bin. The black dashed horizontal line marks the 1-to-1 ratio.}
    \label{fig:noise}
\end{figure}

\section{Applying the method}\label{sec:apply}
We proceeded with the application of \texttt{STACKER} and \texttt{FINDER} to the eighteen MHONGOOSE galaxies with $i\leq60^\circ$ listed in Table \ref{table:obs}. For each galaxy, we have a `single-track' cube obtained from the data reduction of a single 5.5 hours track, and for ten of them we also have a `full-depth' cube obtained by combining all 10 observing sessions.\\\indent We first applied our stacking on the single-track dataset. The reason was to test our algorithm in an `uncontrolled' environment, which means that, unlike when working with simulated data, we are now truly unconstrained, having no information about the priors. Any detection present in the stacked spectra of the single-track cubes was then checked against the corresponding full-depth dataset (when available) to provide an indication on the reliability of each detection and the robustness of any source that will be found in the data.\\\indent Indeed, any source which is present in both the single-track and full-depth stacked spectrum of a given region is likely genuine, while any excess of emission picked up in the single-track cubes but not in the full-depth might provide insight on how to further discriminate plausible emission line from artefacts.

\subsection{Stacking the MHONGOOSE cubes}\label{sec:disc}
As was done for the TNG50 galaxies, we used the \texttt{SoFiA-2} moment map to blank the already detected emission of the target galaxies and their satellites. As discussed in Sect. \ref{sec:assumption}, we employed the systemic velocity as the reference redshift. Again, the regridding has been chosen so that 1 pixel equals 1 beam and the weighting scheme was set to be the equal-weight. This procedure was repeated in the same way for both the single-track and the full-depth data.\\\indent We opt for a common set of parameters for the line finder. To determine this common set we optimised the \texttt{FINDER} parameters for each galaxy following the procedure presented at the end of Sect. \ref{sec:calibration}, that is by minimizing the difference between the observed and expected cumulative distribution of the variable defined by Eq. (\ref{eq:skellam}), and then we took the mean value of each parameter\footnote{Optimizing the values for each galaxy led to negligible differences in the number of detected lines.}. The full list of parameters used for the source finding are given in Table \ref{table:finder}.

\subsection{The Gaussianity of MHONGOOSE noise}\label{sec:noise}
The reliability of any detection extracted by our algorithm is meaningful only if the MHONGOOSE noise distribution is Gaussian. We therefore created a `MHONGOOSE noise cube' by reimaging a full-depth cube in a $\sim1000$ km s$^{-1}$ spectral range away from the target galaxy and free of known emission. The noise level in this cube is $\sigma=0.15$ mJy beam$^{-1}$. To compare with, we also created a cube with the same size but containing values extracted from a Gaussian noise distribution with dispersion $\sigma_0=0.15$ mJy beam$^{-1}$.\\\indent In the left-side panel of Fig. \ref{fig:noise} we plotted the histogram of the flux values ($F$) in the MHONGOOSE noise cube (pink) and compare this with the Gaussian histogram (in blue). Taking the ratio, we directly measured the presence of any non-Gaussianity. In the right-side panel of Fig. \ref{fig:noise} it is shown that the ratio diverges from 1 beyond a level of $\sim4\sigma$ (grey dash-dotted vertical lines).\\\indent For a Gaussian distribution, 0.000063\% of the data have $|F|>4\sigma$, which corresponds to 25606 voxels given the size of the cube. In the MHONGOOSE noise cube we found that instead 0.000069\% of the voxels (27864) have $|F|>4\sigma$. This is equivalent to an excess of 8.8\%.\\\indent Of these 27864 voxels having $|F|>4\sigma$, 14053 are positive ($P$) and 13811 are negative ($N$). The almost symmetric distribution ($\frac{P}{N}=1.02$) suggests that this excess is due to low-level artefacts such as residual continuum emission or RFI. A further confirmation of the above is provided by checking the spatial and spectral location of the voxels with $|F|>4\sigma$. We found that they are spatially randomly distributed and spectrally non-contiguous. \\\indent We also repeated the aforementioned analysis for the Stokes-Q\footnote{Stokes-Q is given by the difference between the two linear polarisations, thus, the signal cancels out and only the noise remains.} component of the MHONGOOSE noise cube, as well as for the single-track and full-depth cubes used for stacking, finding always similar results. Thus, we concluded that the MHONGOOSE noise distribution is mostly Gaussian, with a small excess at the highest absolute flux values. The symmetry of this excess argues that it is inherent to the MHONGOOSE noise distribution, caused by low-level artefacts, rather than real sources.

\begin{table*}[]
    \centering
    \caption{Source finder parameters for each cell size and dataset.}
    \begin{tabular}{c | c c | c c | c c }
        \hline\hline
        & \multicolumn{2}{ c |}{$14\times14$ beams} & \multicolumn{2}{c |}{$9\times9$ beams} & \multicolumn{2}{c}{ $5\times5$ beams} \\ 
        & Single-track & Deep & Single-track & Deep  & Single-track & Deep \\ 
        \hline
        Flux threshold [$\sigma$] & 2.5 & 2.5 & 2.5 & 2.5 & 2.5 & 2.5\\ 
        Smoothing kernels [channels] & 1, 5, 9 & 1, 5, 10 & 1, 7, 10 & 1, 7, 10 & 1, 7, 12 & 1, 7, 12\\ 
        Minimum linewidth [channels] & 5 & 5 & 7 & 7 & 7 & 7\\ 
        Minimum SNR & 5 & 5 & 5 & 5 & 5 & 5\\ 
        Reliability threshold & 0.85 & 0.85 & 0.85 & 0.85 & 0.85 & 0.85\\ 
        \hline
    \end{tabular}
    \label{table:finder}
\end{table*}

\subsection{Detections in the single-track data}\label{sec:single-track}
The stacking on the single-track data cubes of the MHONGOOSE sample results in 25 reliable sources out of 82479 detections, for which the stacked spectra are available at \href{https://zenodo.org/records/14185851}{Zenodo}\footnote{https://zenodo.org/records/14185851} and the main properties are listed in Table \ref{table:single-track}. Three of the authors independently visually inspected the single-track cubes at the spatial and spectral location of each detection and all agreed with the following classifications:
\begin{itemize}
    \item eleven sources were not visually identified. For five we also have the corresponding full-depth data and there was no counterpart there as well. Therefore, these five are likely noise peaks;
    \item ten sources were identified in their single-track cube. However, we were not able to check them also in their full depth cube because it was not available, as the survey is still to be completed at the time of writing;
    \item three sources have been associated with faint H{\sc i} emission from the outer disk of the target galaxies or their companions;
    \item one source is galaxy MKT J125225.4-124304.1, which is listed in \citet{mhongoose2}. In that paper the galaxy is listed with uncertain parameters, as not the entire velocity range was imaged. Also in the stacked spectrum the velocity range is incomplete, however, based on the full-depth line profile the systemic velocity of the galaxy can now be updated to 1279 km s$^{-1}$.
\end{itemize}
These last four sources were detected by \texttt{SoFiA-2} in the corresponding full-depth cube. This detection of H{\sc i} in emission via an unconstrained spectral stacking thus confirms the capability of our procedure.\\\indent We plot the peak, total and mean flux of the 25 sources as a function of their line width and of the cell sizes on the background of Fig. \ref{fig:detections}. We found no preferred size for the stacking regions and no clear threshold in terms of flux properties or line width to discriminate between genuine detections and false positives, although low flux detections are generally not visible by eye in the cube.

\subsection{Detections in the full-depth data}\label{sec:full-depth}
\begin{figure}
    \centering
    \includegraphics[width=\hsize]{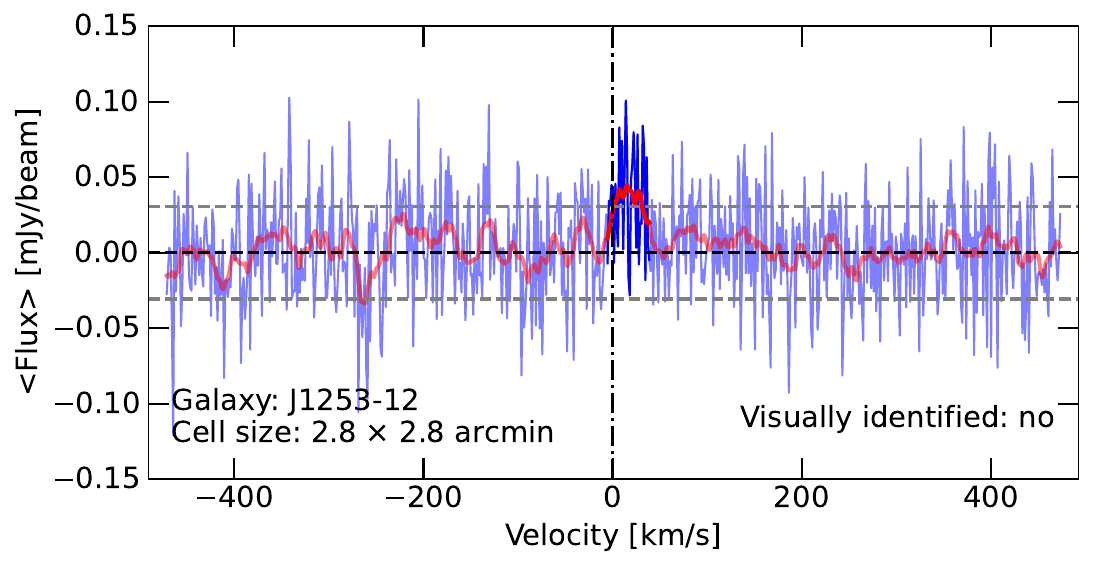}
    \caption{Reliable source around the galaxy J1253-12. The stacked spectrum and its 9-channel boxcar smoothed version are given in blue and red, respectively. The detected source is highlighted. The horizontal grey dashed lines are the $\pm\sigma$ level for the unsmoothed spectrum, while the black dashed line is the 0-flux level. The vertical black dashed-dotted line is, instead, the 0-km s${-1}$ reference velocity. The galaxy name and the cell size are provided in the bottom-left corner, while at the bottom-right is reported if the detection was also visually identified in its corresponding cube..}
    \label{fig:spectrum}
\end{figure}

\begin{figure*}
    \centering
    \includegraphics[width=\hsize]{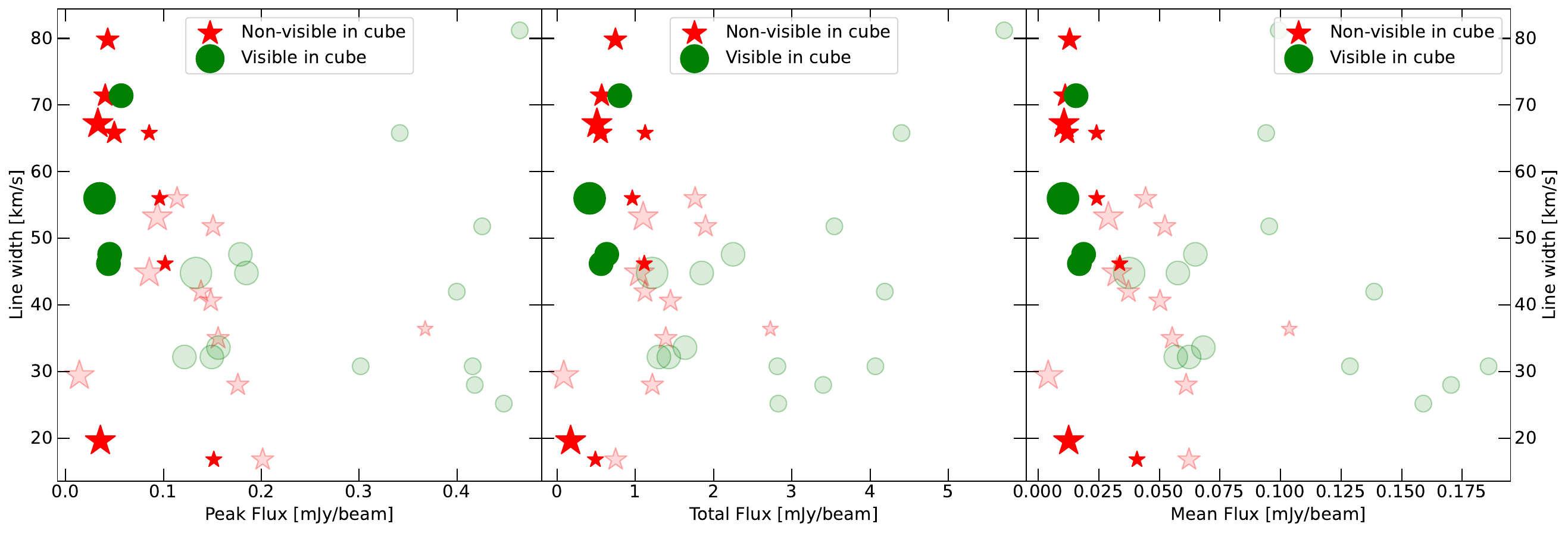}
    \caption{Peak flux (left panel), total flux (central panel) and mean flux (right panel) of the stacking detections as a function of their width. Detections in single-track stacked spectra are provided in the background, while detections in full-depth stacked spectra are given in the foreground. Detections with no counterpart visible in the cube are labelled with red stars. Green circles refer to visually identified H{\sc i} emission. The size of the markers is indicative of the size of the square regions used for the stacking.}
    \label{fig:detections}
\end{figure*}

The stacking on the full-depth data leads to a total of thirteen reliable sources out of 34839 detections corresponding to a detection rate of $\sim2.1$ sources per cube, slightly higher than for the single-track data ($\sim1.8$). The full source list is provided in Table \ref{table:full-depth} and a representative stacked spectrum for a reliable detection is given in Fig. \ref{fig:spectrum}. The full sample of reliable detections is available online at \href{https://zenodo.org/records/14185851}{Zenodo}\footnote{https://zenodo.org/records/14185851}. By visually inspecting the cubes we found that four sources are identifiable by eye, while nine do not have a clear counterpart in the cube. These numbers are equivalent to $\sim15\%$ of the reliable detections being visible by eye, $\sim15\%$ not unambiguously distinguishable and $\sim70\%$ not visible. For the single-track data, these percentage are $\sim20\%$, $\sim36\%$ and $\sim44\%$, respectively.\\\indent The larger number of non-visible detections is due to the flux properties of the sources found in the stacked spectra of the full-depth cubes. The comparison with single-track detections made in Fig. \ref{fig:detections} shows that the full-depth detections lie in the region of parameter space where also for the single-track observations we have not found the corresponding source in the H{\sc i} cube. Given the analysis of the MeerKAT noise presented in Sect. \ref{sec:noise}, we suspect the non-identified sources are not low-level artefacts still present in the cubes but rather genuine H{\sc i} excess, although only through a direct detection via deeper observations we will be able to confirm this.

\subsection{Explaining the missing gas}\label{sec:explain}
In Sect. \ref{sec:calibration} we have presented the application of our algorithm to two TNG50 mock galaxies detecting emission outside the \texttt{SoFiA-2} mask. However, we find very few plausible H{\sc i} detections when stacking the MHONGOOSE observations. We suspect that the reason why little excess of H{\sc i} is found in the full-depth cubes is that the radial profile of the H{\sc i} disks exhibits a sharp decrease in the column density, because the gas likely becomes ionized, thus requiring a detection limit well below $10^{17}$ cm$^{-2}$ in order to be detected. This has been suggested previously to occur at higher column densities as well \citep{maloney93,dove94,ianjam18}, and a radial profile study of 18 MHONGOOSE galaxies made with the Green Bank Telescope (GBT) seem to support the view of this decrease happening at the $\sim10^{17}$ cm$^{-2}$ column density level \citep{sardone21}.\\\indent Recent observations with the GBT of NGC 891 and NGC 4565 did also reveal that the H{\sc i} column density in the CGM/IGM is $<10^{17}$ cm$^{-2}$ \citep{das24}. Similar results were achieved with the Five-Hundred-meter Aperture Spherical radio Telescope by \citet{xu22}, observing of Stephan's Quintet, and \citet{liu23} pointing at NGC 4490/85. Note that, as discussed in Sect. \ref{sec:intro}, single-dish observations are limited by their low spatial resolution ($>200''$). Reaching the $<10^{17}$ cm$^{-2}$ sensitivity at a higher angular resolution with interferometers will require significant observing time even with the next generation of interferometers (such as SKA-MID, see Fig. 2 of \citealt{mhongoose2}).

\section{Conclusion}\label{sec:conc}
We present a method to detect emission from the H{\sc i} in the CGM/IGM around resolved nearby galaxies with no prior information about the spatial and spectral distribution of the gas. The method consists of a standard stacking procedure, described in Sect. \ref{sec:stacking}, combined with a one-dimensional line finder which uses the reliability calculation presented in \citet{serra12} to discriminate noise peaks from potentially genuine emission (see Sect. \ref{sec:finder}).\\\indent We applied our algorithm to two TNG50 mock galaxies taken from the \citet{ramesh23} sample and optimized its free parameters, mainly, the size of the stacking regions, the weighting scheme and the source finder parameters (Sect. \ref{sec:calibration}).\\\indent As simulated data provide us with exact knowledge of the gas kinematics, we tested whether using the systemic velocity or the rotation velocity to align the spectra provides the best result (Sect. \ref{sec:assumption}) and found that for averaged-inclined galaxies using the systemic velocity is preferable.\\\indent The TNG50 data has also been used to study the limitations of the stacking, as discussed in Sect. \ref{sec:limit}, finding that imposing a minimum SNR in the stacked spectrum leads inevitably to the loss of the majority of the detections whose peak flux falls above the detection limit (see Fig. \ref{fig:snrmaps}). However, lowering the SNR threshold to include more signal is not a viable option, as this will also increase the number of false detections.\\\indent TNG50 data are, by construction, filled with Gaussian noise. In Sect. \ref{sec:noise} we quantified the level of Gaussianity of the MHONGOOSE noise, finding a $\sim10\%$ excess of voxels with absolute flux $>4\sigma$ compared to a Gaussian.\\\indent When stacking the MHONGOOSE data (Sect. \ref{sec:full-depth}), we find very little excess H{\sc i} emission. As discussed in Sect. \ref{sec:explain}, assuming our stacking method works, the lack of H{\sc i} at the $\sim10^{17}$ cm$^{-2}$ level beyond the periphery of the H{\sc i} disks may be the result of a sudden drop in the H{\sc i} column density of the gaseous disk, as radial profile studies using GBT observations are suggesting \citep{sardone21}. If validated by future analysis of the H{\sc i} radial profiles of the MHONGOOSE galaxies with the higher resolution MeerKAT data, it will confirm the need for further comparisons with simulations where extended H{\sc i} distributions are a more general feature.\\\indent Stacking also has its limitations. Increasing the number of stacked spectra does not necessarily mean that the SNR of the emission we are trying to detect will increase. However, if our conclusions hold, then longer integration times can help with the direct detection of the H{\sc i}. One would need many tens of hours even on SKA-MID to reach the necessary column density limit and be able to resolve features even at arc-minute resolution.

\section{Future Prospects}\label{sec:future}
This work represents a first step in the application of unconstrained spectral stacking to detect H{\sc i} in emission around nearby resolved objects. There is still space for further improvements. For example, future studies can focus on designing a more sophisticated way to minimize the discrepancy between the assumed kinematics and the real velocity of the gas. This can be done, for instance, by repeating the analysis presented in Sect. \ref{sec:assumption} on a much larger sample of simulated galaxies and study which is the best assumption as a function of the inclination angle. Alternatively, one can derive the rotation curve and model the velocity field leaving the kinematics and geometry of the outer disk as free parameters, to check which configuration maximizes the number of detections. The work by \citet{das24} is a step in this direction, as they stacked along the major and minor axis where it is simpler to determine the kinematics of the gas, but ultimately one wants to detect the H{\sc i} all around the galaxies.\\\indent Another limitation of our method is the use of regions with fixed shape and size. Implementing a dynamical determination of the stacking regions where the shape and size of the stacking regions is continuously varied until a line is detected and its SNR maximized, like employing Voronoi tessellation \citep{voronoi1,voronoi2}, may lead to even more reliable identification of any excess of H{\sc i} embedded into the noise.\\\indent Another way forward to understand the properties of the H{\sc i} in the CGM of nearby isolated star-forming galaxies may consist in the application of this stacking procedure on the whole MHONGOOSE sample normalized in size and redshift. This experiment will resemble the `typical' stacking analysis carried out to study the mean properties of sample of galaxies (e.g., \citealt{fabello11,delhaize13,gereb14,gereb15a,gereb15b,kanekar16,maccagni17,ianjam18,hiss,healy21c,chowdhury22,amiri23,apurba23}).\\\indent As mentioned in the previous section, it is imperative to do a detailed radial profile study of the H{\sc i} in the MHONGOOSE galaxies which will improve the results of \citet{sardone21}, further constraining the shape and extension of the gaseous disk.\\\indent Finally, it is also worth noting that as the limit on the H{\sc i} column density outside the gaseous disk keeps getting lower, it will be possible to better distinguish between various modes of accretion (for example, cold vs.\ hot accretion). Further work on the MHONGOOSE data, but also future deep surveys should be able to provide more observational constraints on these scenarios.

\begin{acknowledgements}
We thank the anonymous referee for the constructive comments. This work has received funding from the European Research Council (ERC) under the European Union’s Horizon 2020 research and innovation programme (grant agreement No 882793 `MeerGas'). SV thanks N. Kanekar for the insightful discussion about the stacking technique. KS acknowledges support from the Natural Sciences and Engineering Research Council of Canada (NSERC). NZ and DJP gratefully acknowledge support by the South African Research Chairs Initiative of the Department of Science and Technology and the National Research Foundation. PK is partially supported by the BMBF project 05A23PC1 for D-MeerKAT. L.C. acknowledges financial support from the Chilean Agencia Nacional de Investigacion y Desarrollo (ANID) through Fondo Nacional de Desarrollo Cientifico y Tecnologico (FONDECYT) Regular Project 1210992.
\end{acknowledgements}

\bibliographystyle{aa}
\bibliography{reference.bib}

\begin{appendix}
\onecolumn
\section{Stacking detections}
\begin{table*}[h!]
    \centering
    \caption{The used MHONGOOSE data cubes and the detections found in the stacked spectra.}
    \begin{tabular}{c | c | c | c | c}
        \hline\hline
        Galaxy & Single-track & Detected sources & Deep & Detected sources \\ 
        & $r=1.5$, $t=0''$ & Reliable / Non-reliable & $r=1.5$, $t=0''$ & Reliable / Non-reliable\\ 
        \hline
            J0008-34 & X\tablefootmark{a} & - & N.A. & - \\ 
            J0031-22 & \checkmark & 2 / 5099 & N.A. & - \\ 
            J0052-31 & \checkmark & 1 / 6042 & N.A. & - \\  
            J0309-41 & X\tablefootmark{b} & - & X\tablefootmark{b} & - \\ 
            J0310-39 & \checkmark & 0 / 5806 & N.A. & - \\ 
            J0335-24 & \checkmark & 2 / 6206 & X\tablefootmark{a} & - \\ 
            J0351-38 & \checkmark & 2 / 6135 & N.A. & - \\ 
            J0419-54 & \checkmark & 1 / 6011 & \checkmark & 2 / 6029 \\ 
            J0429-27 & \checkmark & 1 / 6124 & \checkmark & 0 / 5957 \\ 
            J0445-59 & \checkmark & 2 / 6047 & N.A. & - \\ 
            J0454-53 & \checkmark & 6 / 5774 & N.A. & - \\ 
            J0459-26 & \checkmark & 2 / 6003 & X\tablefootmark{a} & - \\ 
            J0546-52 & \checkmark & 0 / 6248 & \checkmark & 2 / 6016 \\  
            J1253-12 & \checkmark & 1 / 5905 & \checkmark & 1 / 5937 \\  
            J1318-21 & \checkmark & 2 / 5765 & \checkmark & 5 / 5651 \\  
            J1337-28 & \checkmark & 1 / 5314 & \checkmark & 3 / 5249 \\ 
            J2257-41 & X\tablefootmark{a} & - & X\tablefootmark{a} & - \\ 
            J2357-32 & X\tablefootmark{c} & - & N.A. & - \\ 
        \hline
    \end{tabular}
    \tablefoot{N.A. $=$ Not Available.\\ 
    \tablefoottext{a}{Excluded due to residuals as a result of the deconvolution;} \tablefoottext{b}{Excluded due to non-optimal noise distribution;} \tablefoottext{c}{Excluded due to non-filtered diffuse Magellanic Stream emission.}} 
    \label{table:obs}
\end{table*}

\begin{table*}[h!]
    \centering
    \caption{Stacking detection in the single-track cubes.}
    \begin{tabular}{c | c | c | c | c | c | c | c}
    \hline\hline
        Galaxy & Cell $\alpha$ & Cell $\delta$ & Cell size & $V_0$ & Line width & $F_{tot}$ & Visually identified\\ 
        & [hh:mm:ss] & [deg:mm:ss] & [arcmin] & [km s$^{-1}$] & [km s$^{-1}$] & [mJy beam$^{-1}$] & \\ 
        \hline
        J0031-22 & 00:30:32 & -23:09:36 & 5.8 & 667 & 16.8 & 0.7 & No\\
        J0031-22 & 00:29:18 & -22:57:02 & 5.8 & 722 & 42.0 & 1.1 & No\\
        J0052-31 & 00:51:55 & -31:13:59 & 2.8 & 1571 & 81.2 & 5.6 & Yes\tablefootmark{a} \\  
        J0335-24 & 03:37:42 & -24:28:24 & 2.8 & 1532 & 51.8 & 3.5 & Yes\tablefootmark{b} \\
        J0335-24 & 03:37:37 & -24:33:02 & 5.8 & 1485 & 47.6 & 2.2 & Yes\tablefootmark{b} \\
        J0351-38 & 03:52:04 & -38:14:13 & 8.7 & 974 & 29.4 & 0.8 & No \\
        J0351-38 & 03:49:44 & -38:33:22 & 5.8 & 845 & 56.0 & 1.8 & No \\
        J0419-54 & 04:17:10 & -54:43:08 & 2.8 & 1512 & 36.4 & 2.7 & No \\  
        J0429-27 & 04:32:45 & -27:42:04 & 5.8 & 1001 & 35.0 & 1.4 & No \\  
        J0445-59 & 04:47:59 & -58:51:39 & 5.8 & 1290 & 44.8 & 1.8 & Yes\tablefootmark{c} \\  
        J0445-59 & 04:40:26 & -59:02:59 & 5.8 & 1341 & 51.8 & 1.9 & No \\ 
        J0454-53 & 04:53:45 & -53:26:14 & 8.7 & 323 & 44.8 & 1.2 & Yes\tablefootmark{c} \\  
        J0454-53 & 04:53:35 & -53:27:49 & 5.8 & 328 & 33.6 & 1.6 & Yes\tablefootmark{c} \\  
        J0454-53 & 04:52:18 & -53:04:18 & 5.8 & 346 & 30.8 & 1.3 & Yes\tablefootmark{c} \\  
        J0454-53 & 04:54:05 & -53:34:57 & 2.8 & 321 & 30.8 & 2.8 & Yes\tablefootmark{c} \\  
        J0454-53 & 04:53:45 & -53:32:02 & 2.8 & 326 & 28.0 & 3.4 & Yes\tablefootmark{c} \\  
        J0454-53 & 04:53:45 & -53:29:07 & 2.8 & 328 & 25.2 & 2.8 & Yes\tablefootmark{c} \\  
        J0454-53 & 04:52:27 & -53:05:45 & 2.8 & 344 & 42.0 & 4.1 & Yes\tablefootmark{c} \\  
        J0459-26 & 05:01:42 & -25:43:59 & 5.8 & 851 & 40.6 & 1.4 & No \\  
        J0459-26 & 04:59:32 & -26:30:42 & 5.8 & 818 & 32.2 & 1.4 & Yes\tablefootmark{c} \\  
        J1253-12 & 12:55:39 & -12:05:05 & 2.8 & 916 & 65.8 & 4.3 & Yes\tablefootmark{b} \\  
        J1253-12 & 12:52:28 & -12:42:59 & 2.8 & 1284 & 29.4 & 3.3 & Yes\tablefootmark{b} \\   
        J1318-21 & 13:20:29 & -21:06:55 & 8.7 & 690 & 53.2 & 1.1 & No \\  
        J1318-21 & 13:18:30 & -21:31:43 & 5.8 & 716 & 29.4 & 1.2 & No \\  
        J1337-28 & 13:35:01 & -28:42:11 & 8.7 & 489 & 44.8 & 1.0 & No \\  
        \hline
    \end{tabular}
    \tablefoot{Cell $\alpha$ and cell $\delta$ are the central right ascension and declination of the cell. $V_0$ is the central velocity of the line (in the non-shuffled cube), and $F_{tot}$ is its total flux.\\ 
    \tablefoottext{a}{Clear detection. Full-depth cube not available;} \tablefoottext{b}{Confirmed in the full-depth cube;} \tablefoottext{c}{Plausible detection. Full-depth cube not available.}}
    \label{table:single-track}
\end{table*}

\begin{table*}[h!]
    \centering
    \caption{Stacking detection in the full-depth cubes.}
    \begin{tabular}{c | c | c | c | c | c | c | c}
    \hline\hline
        Galaxy & Cell $\alpha$ & Cell $\delta$ & Cell size & $V_0$ & Line width & $F_{tot}$ & Visually identified\\ 
        & [hh:mm:ss] & [deg:mm:ss] & [arcmin] & [km s$^{-1}$] & [km s$^{-1}$] & [mJy beam$^{-1}$] & \\ 
        \hline
        J0419-54 & 04:23:34 & -54:51:57 & 8.7 & 1500 & 56.0 & 0.4 & Yes \\ 
        J0419-54 & 04:15:23 & -55:35:29 & 8.7 & 1475 & 67.2 & 0.5 & No \\  
        J0546-52 & 05:49:37 & -52:28:37 & 5.8 & 1158 & 47.6 & 0.6 & Yes \\ 
        J0546-52 & 05:42:02 & -51:30:11 & 5.8 & 1253 & 79.8 & 0.7 & No \\ 
        J1253-12 & 12:53:28 & -12:05:03 & 2.8 & 837 & 46.2 & 1.1 & No \\  
        J1318-21 & 13:19:52 & -20:58:10 & 8.7 & 523 & 19.6 & 0.2 & No \\
        J1318-21 & 13:19:20 & -21:08:24 & 5.8 & 704 & 71.4 & 0.8 & Yes\tablefootmark{a} \\  
        J1318-21 & 13:18:05 & -20:56:43 & 5.8 & 634 & 46.2 & 0.6 & Yes\tablefootmark{a} \\  
        J1318-21 & 13:17:40 & -20:44:56 & 5.8 & 678 & 65.8 & 0.6 & No \\
        J1318-21 & 13:16:01 & -20:38:56 & 5.8 & 681 & 71.4 & 0.6 & No \\
        J1337-28 & 13:39:53 & -28:13:00 & 2.8 & 930 & 16.8 & 0.5 & No \\ 
        J1337-28 & 13:38:53 & -27:46:51 & 2.8 & 492 & 56.0 & 1.0 & No \\ 
        %J1337-28 & 13:35:29 & -27:46:48 & 2.8 & 526 & 25.2 & 0.7 & No \\  
        J1337-28 & 13:34:48 & -28:18:52 & 2.8 & 573 & 65.8 & 1.1 & No \\
        \hline
    \end{tabular}
    \tablefoot{Cell $\alpha$ and cell $\delta$ are the central right ascension and declination of the cell. $V_0$ is the central velocity of the line (in the non-shuffled cube), and $F_{tot}$ is its total flux.\\ 
    \tablefoottext{a}{Plausible detection.}}
    \label{table:full-depth}
\end{table*}
\clearpage

\twocolumn
\section{Calibration on TNG50 555013}\label{sec:app1}
\begin{figure}[!h]
    \centering
    \includegraphics[width=\hsize]{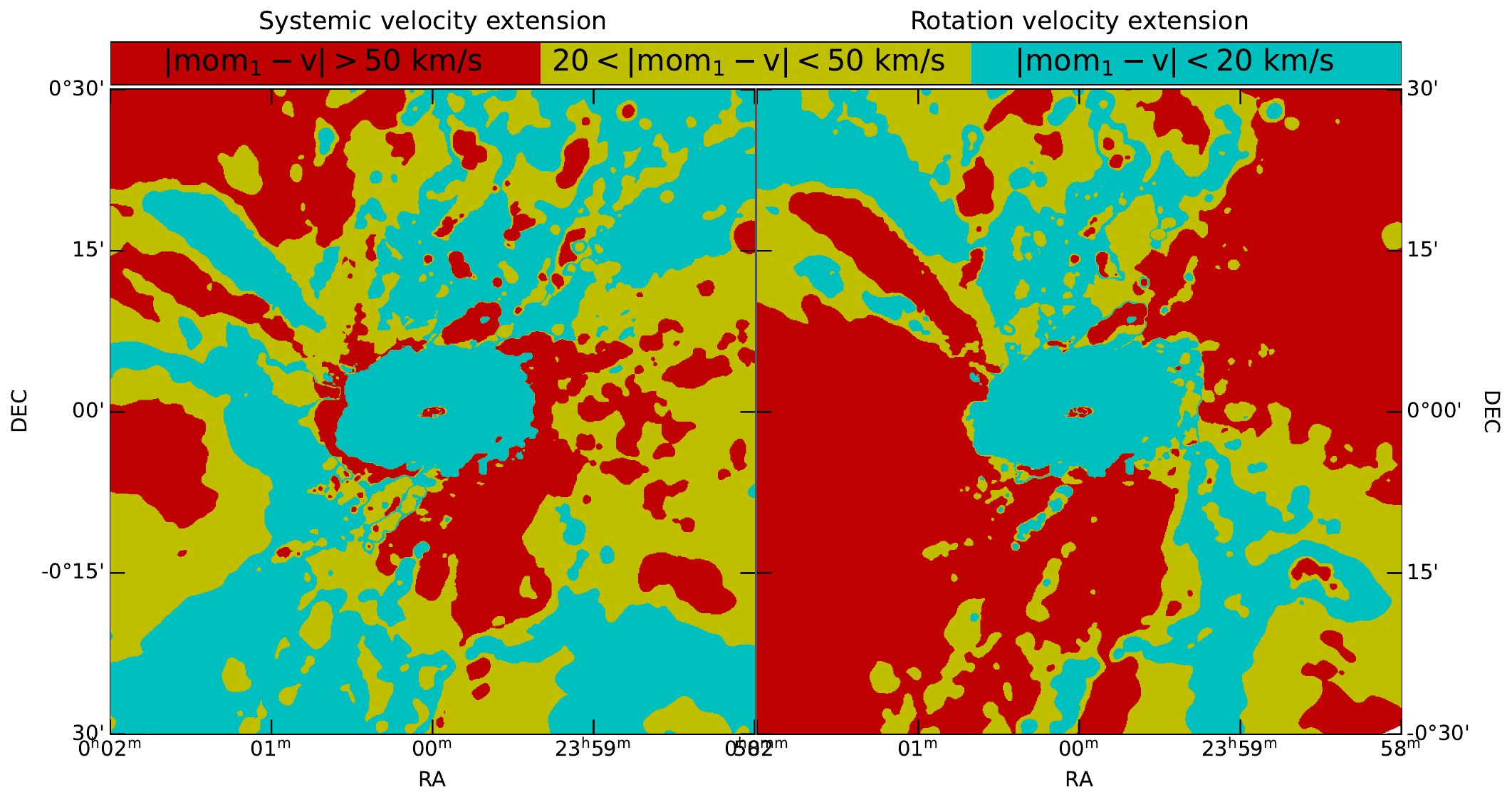}
    \caption{Error in the spectral alignment for the TNG50 galaxy 555013 under two different CGM/IGM kinematics assumptions. \textit{Left panel}: the residuals in terms of $mom_1-v$, where $\text{mom}_1$ is the noiseless moment 1 map and $v$ the assumed kinematics, when $v=v_{sys}$. Light-blue corresponds to $|\text{mom}_1-v|<20$ km s$^{-1}$, yellow to $20<|\text{mom}_1-v|<50$ km s$^{-1}$ and red is $|\text{mom}_1-v|>50$ km s$^{-1}$. \textit{Right panel}: same as left panel but when $v$ is given by assuming the gas is co-rotating with the galaxy with a flat rotation curve.}
    \label{fig:555kin}
\end{figure}

\begin{figure}
    \centering
    \includegraphics[width=\hsize]{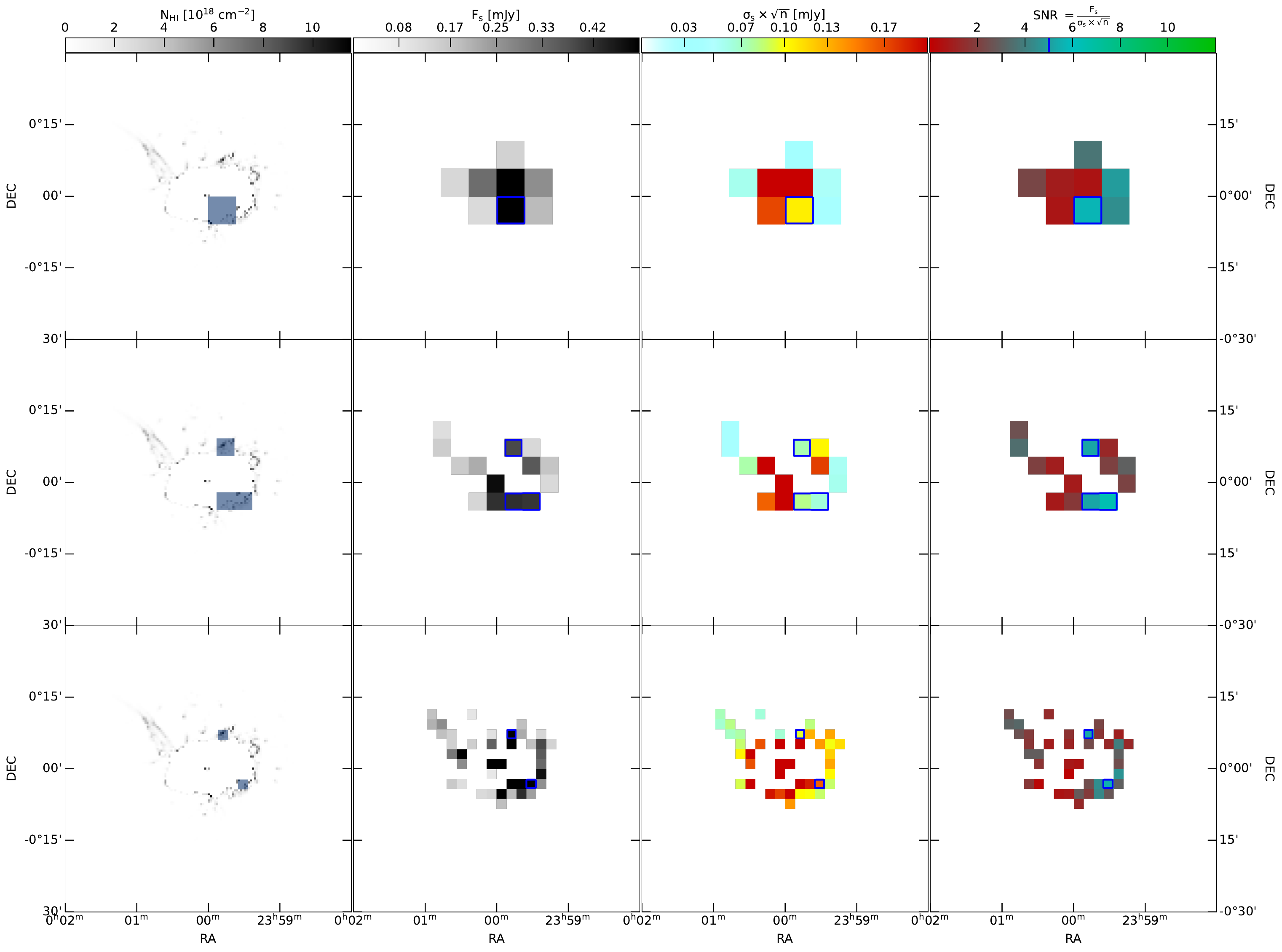}
    \caption{Stacking SNR maps for the TNG50 galaxy 55013. \textit{First column}: stacking regions where SNR $>5$ (blue squares) overlaid with the grey-scale noiseless moment 0 map, blanked from the emission already detected by \texttt{SoFiA-2}. Top to bottom is for cell sizes of $14\times14$, $9\times9$ and $5\times5$ beams, respectively. \textit{Second column}: the integrated flux of the signal for stacking regions where the H{\sc i} column density is $>3.6\times10^{17}$ cm$^{-2}$. The blue contours enclose the cells where SNR $>5$. \textit{Third column}: the noise of the stacked spectrum multiplied by the square root of the number of channels covered by the source. \textit{Fourth column}: integrated SNR of the signal in the stacked spectrum for the regions. The SNR $=5$ level is indicated also on the colourbar.}
    \label{fig:555snr}
\end{figure}

In Sect. \ref{sec:assumption} and Sect. \ref{sec:limit} we presented the results of testing our method on the TNG50 galaxy 520885. For completeness, we report here also the outcomes for the TNG50 galaxy 555013. This mock galaxy is mainly characterized by a regular disk and an extended tail in the north-east. In Fig. \ref{fig:555kin} we report the alignment residuals in terms of $|mom_1-v|$, where $v$ is the assumed kinematics. Also for this galaxy most of the residuals are $<20$ km/s when the systemic velocity is used to describe the gas kinematics. Finally, in Fig. \ref{fig:555snr} it is shown that the SNR of the stacked spectra is lower than the threshold of 5 for almost all the cells, likely due to the overall lower column density of the gas outside the disk with respect to TNG50 520885.

\clearpage

\section{The effect of regridding}
\begin{figure}[!h]
    \centering
    \includegraphics[width=\hsize]{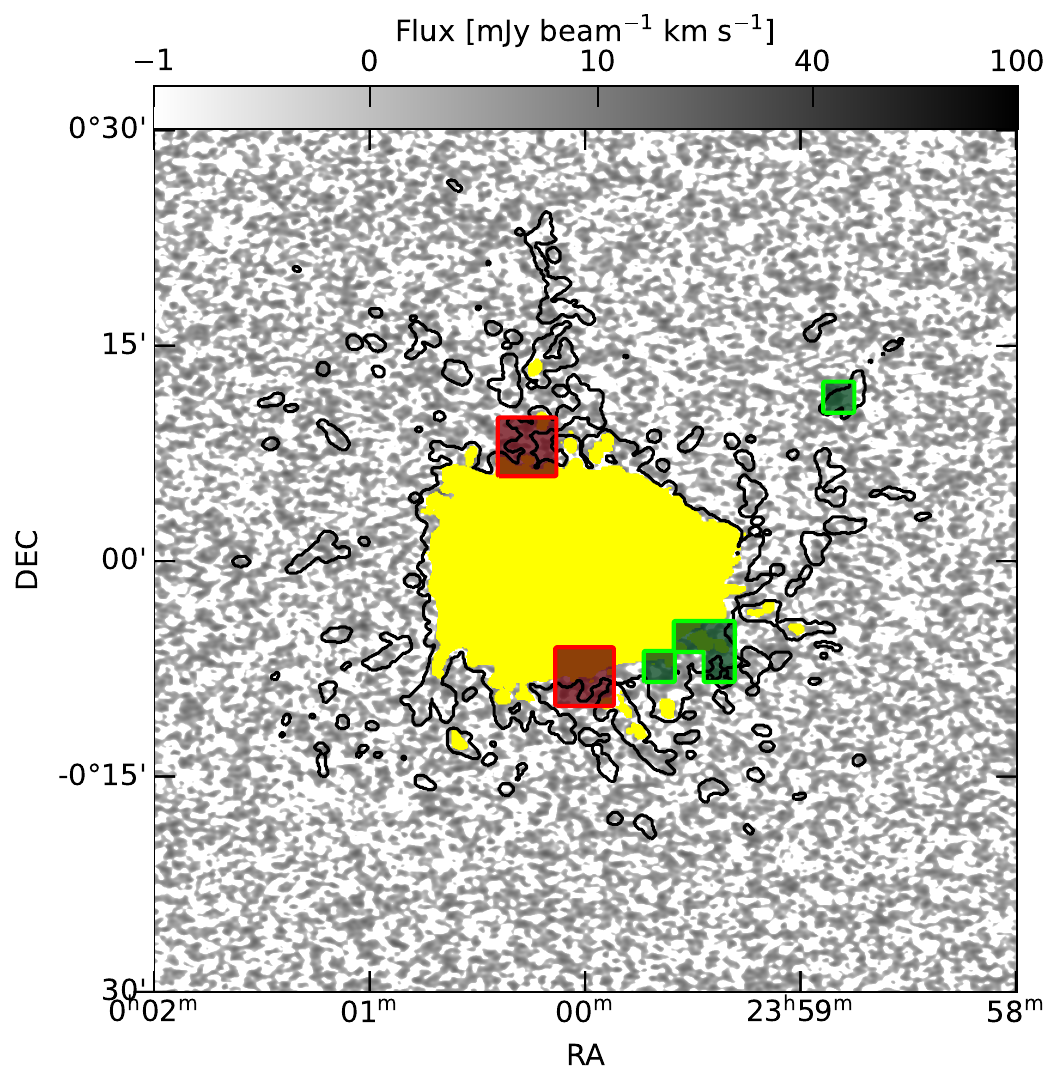}
    \caption{Stacking detection map without regridding the cube. The background grey-scale image is the cube of the TNG50 520855 galaxy collapsed along the spectral axis and blanked from the galaxy emission. Black contours denote the noiseless moment 0 map clipped at the column density value of $3.6\times10^{17}$ cm$^{-2}$. The coloured squares indicate the stacking regions where their stacked spectrum contains a reliable detection. Different colours represent different cell size: red for $9\times9$ and green for $5\times5$ beams.}
    \label{fig:calnoregrid}
\end{figure}

One of the preliminary steps of our stacking procedure is to regrid the cube such that the pixel size is upscaled to match the beam size. One potential effect of the regridding is to lose flux and thus detections. The argument is that after regridding the cube violates the Nyquist sampling criterion, as now the pixel size does not match the longest baseline of the interferometer, implying information is lost. To check how important this may be for our stacking experiment, we apply the procedure presented in Sect. \ref{sec:calibration} without regridding.\\\indent In Fig. \ref{fig:calnoregrid} we show the same overlays of Fig. \ref{fig:cal}. The comparison between the two figures is showing that without regridding we pick up less sources. We think that the reason for this counter-intuitive conclusion is that the features we aim to pick up via stacking, i.e., H{\sc i} clouds, are sufficiently large that even after regridding they span more than 2.2 beams, satisfying so the Nyquist sampling.

\section{Stacking is not equivalent to smoothing}
\begin{figure}[!h]
    \centering
    \includegraphics[width=\hsize]{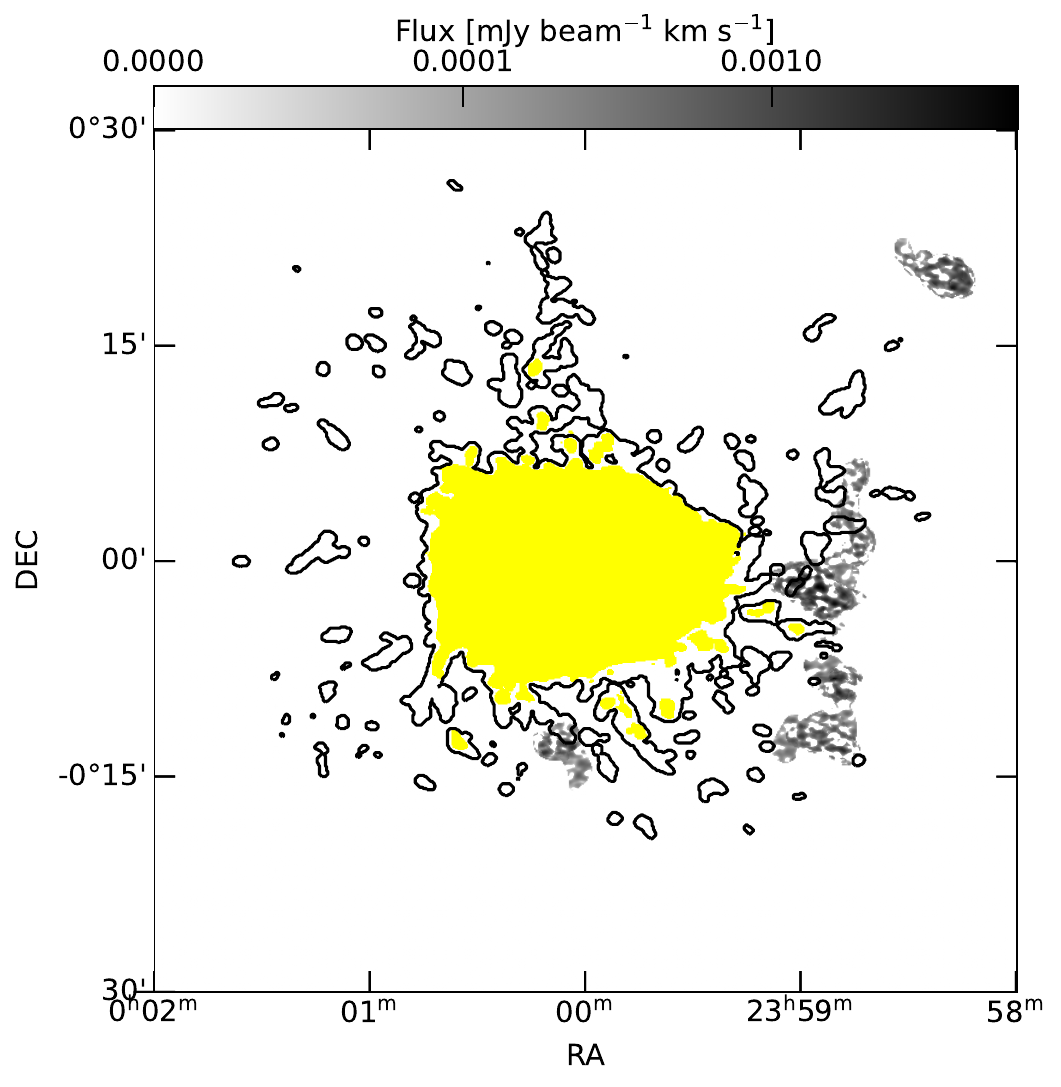}
    \caption{Moment 0 map, blanked from the galaxy emission in yellow, produced by \texttt{SoFiA-2} when emulating our stacking procedure via smoothing. Black contours denote the noiseless moment 0 map clipped at the column density value of $3.6\times10^{17}$ cm$^{-2}$.}
    \label{fig:smoothcal}
\end{figure}

One can argue that the stacking procedure presented in this paper is equivalent to spectrally and spatially smoothing the data at different scales. To check whether this is true we run \texttt{SoFiA-2} on the masked TNG50 520855 galaxy cube using a spatial smoothing kernel equal to our smallest cell size (i.e., $25\times25$ pixels, equivalent to $5\times5$ beams) and spectral smoothing kernels equal to what we have employed with our \texttt{FINDER} (i.e., 1, 7 and 12 channels. cf Table \ref{table:tngfinder}). The reliability parameters (minimum SNR of 5 and reliability threshold of 0.85) and the minimum allowed spatial and spectral sizes for a source ($1\times1$ beam and 7 channels) were set to be equivalent to what we used in our stacking experiment.\\\indent In Fig. \ref{fig:smoothcal} we show the moment 0 map as produced by \texttt{SoFiA-2}. Comparing it with Fig. \ref{fig:cal}, one can see that our stacking procedure is not equivalent to smoothing the cube at different scale. Furthermore, given that the \texttt{SoFiA-2} detections are outside the $3.6\times10^{17}$ cm$^{-2}$ column density level, we suspect that what was detected by \texttt{SoFiA-2} is mainly noise. We want to stress that this comparison is meant to show the differences between our stacking and smoothing and not to compare our procedure with \texttt{SoFiA-2}.

\clearpage

\end{appendix}

\end{document}